\documentclass[journal]{IEEEtran}
\usepackage{amsmath}
\usepackage{amsfonts,amssymb}
\usepackage{graphicx}
\allowdisplaybreaks[4]
\usepackage{xcolor}
\definecolor{deepred}{RGB}{205,38,38}
\usepackage{algorithm}
\usepackage{algorithmic}
\usepackage{amsthm}

\usepackage{lineno,hyperref}
\modulolinenumbers[5]
\usepackage{bm}

\usepackage{multirow}
\usepackage{array}
\usepackage{booktabs}
\usepackage{lipsum}
\usepackage{xcolor}
\usepackage{etoolbox}
\newcommand\norm[1]{\left\lVert#1\right\rVert}

\ifCLASSINFOpdf

\else

\fi

\ifCLASSOPTIONcompsoc
  \usepackage[caption=false,font=normalsize,labelfont=sf,textfont=sf]{subfig}
\else
  \usepackage[caption=false,font=footnotesize]{subfig}
\fi

\begin{document}
\bibliographystyle{ieeetr} 

\title{Optimal Power Flow for Integrated Primary-Secondary Distribution Networks with Center-Tapped Service Transformers}
\author{
\IEEEauthorblockN{Rui Cheng, \textit{ Member}, \textit{IEEE}, 
Naihao Shi, \textit{Student Member}, \textit{IEEE}, 
Zhaoyu Wang, \textit{Senior Member}, \textit{IEEE}, Zixiao Ma, \textit{ Member},  \textit{IEEE}
}
\thanks{}
}
\maketitle

\begin{abstract}
Secondary distribution networks (SDNets) play an increasingly important role in smart grids due to a high proliferation of distributed energy resources (DERs) in SDNets. However, most existing optimal power flow (OPF) problems do not take into account SDNets with center-tapped service transformers. Handling the nonlinear and nonconvex SDNet power flow constraints is still an outstanding problem. To meet this gap, we first utilize the second-order cone programming relaxation and linearization  to make center-tapped  service transformer constraints convex, respectively. Then,  a linearized triplex service line power flow model, including its compact matrix-vector form, is further developed to compose the SDNet OPF model with our proposed center-tapped service transformer model. 
This proposed SDNet OPF model can be easily embedded into existing primary distribution network (PDNet) OPF models, resulting in a holistic power system decision-making solution for integrated primary-secondary distribution networks.  Case studies are presented for two different integrated primary-secondary distribution networks that demonstrate the effectiveness and superiority of this model.
\end{abstract}

\begin{IEEEkeywords}
Secondary distribution network, center-tapped service transformer, integrated primary-secondary distribution networks, optimal power flow (OPF).
\end{IEEEkeywords}

\section{Introduction}
Optimal power flow (OPF) is a powerful tool that is widely used in the operation and planning of power systems to optimize economic, reliability, and environmental objectives.
Nowadays, OPF at the distribution network level is increasingly compelling due to the high penetration of distributed energy resources (DERs) in distribution networks. One of the biggest challenges for OPF at the distribution network level is the nonlinear power flow equations make OPF non-convex, hindering the high-efficiency computation and calculation. The convex relaxation and linearization for OPF at the distribution network level  have been widely studied and investigated in recent decades.

Most studies on the convex relaxation and linearization for OPF at the distribution network level are for primary distribution networks (PDNets), which are typically medium-voltage networks. It can be further classified into two categories: (1)  single-phase PDNet OPF; (2) multi-phase unbalanced PDNet OPF.

For single-phase  PDNets,
second-order cone programming (SOCP) relaxations \cite{RAJ}-\cite{BFM}, semidefinite (SDP) relaxations \cite{XB}-\cite{JL}, and chordal relaxations \cite{MAS}-\cite{SB} are explored to convexify its OPF problems. The {linearization} of non-linear single-phase power flow, based on the bus injection model, is proposed in \cite{SVD}. The classical linearized distribution flow (\textit{LinDistFlow}) \cite{MB}-\cite{MB2}, based on the branch flow model, is developed by neglecting the non-linear power loss term. Moreover, the compact representation of LinDistFlow is proposed in \cite{MF}-\cite{VK} by  means of graph-based matrices. However, 
those works on PDNet OPF cannot be applied to  the real-world multi-phase distribution  networks. To increase empirical fidelity, researchers and engineers strive to explore the convex relaxation and {linearization}  for multi-phase PDNet OPF. SDP relaxations are proposed for unbalanced microgrids in \cite{EDAHZ} and radial unbalanced PDNets in \cite{LG}, respectively. The LinDistFlow model is further extended to multi-phase unbalanced PDNets in \cite{LG}-\cite{BAR}. Some advanced online linearization methods are studied to make the unbalanced PDNet power flow linear. For example, an online feedback-based linearized power flow model  for unbalanced PDNets \cite{OnlinePF} is studied by leveraging the instantaneous voltage and load measurements to update its parameters in real time. 
In addition, {some researchers in \cite{businj_1}-\cite{businj_3} consider transformers in the PDNet OPF by simply expressing and including them as the PDNet admittance matrix.} 


\begin{figure*}[thb]
    \centering
    \includegraphics[width=7.5in]{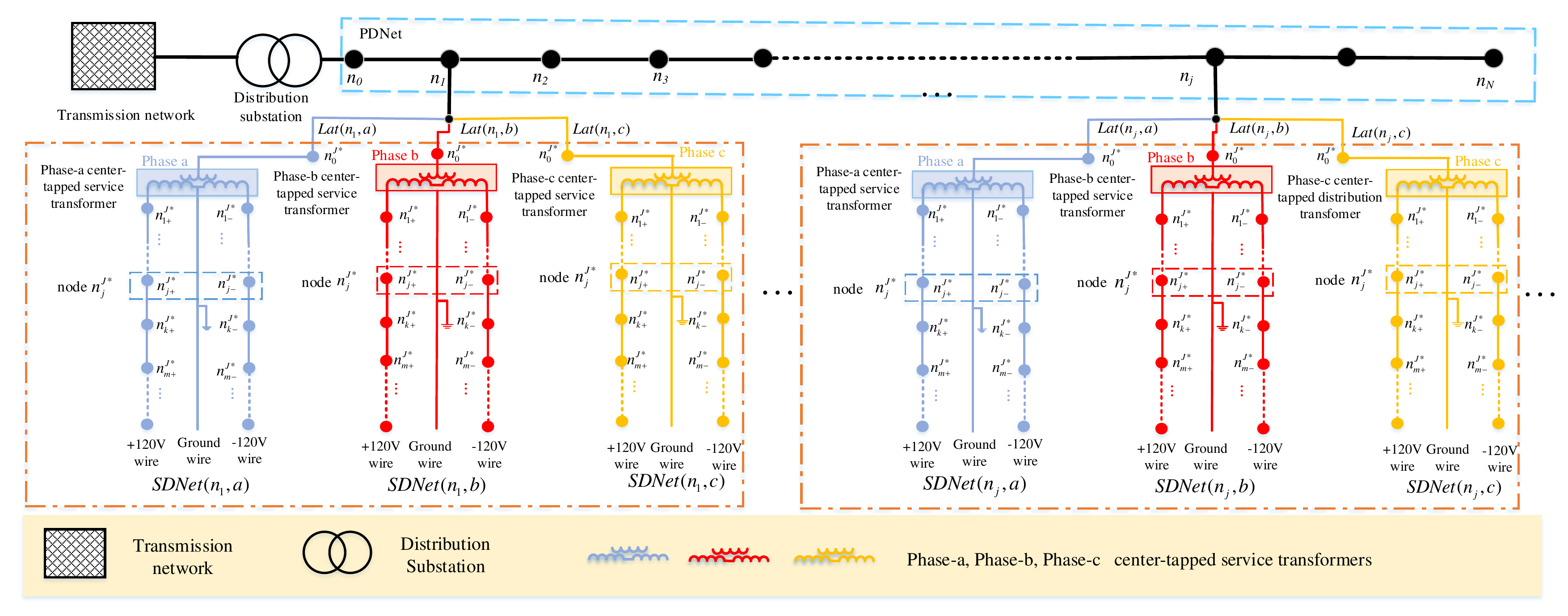}
    \caption{{Integrated Primary-Secondary Distribution Network Illustration}}
    \label{fig:TPSNet}
\end{figure*}

However, all these works focus on single-phase and multi-phase PDNets without considering low-voltage secondary distribution networks (SDNets). Meanwhile, recent years have witnessed the increasing proliferation of loads and DERs connected to {SDNets}. In previous distribution network OPF works, loads and DERs in SDNets are aggregated at the PDNet bus, neglecting the power flow in \textit{SDNets with service transformers}. In U.S., loads and DERs in SDNets are connected to the PDNet via center-tapped service transformers. Such an aggregation could introduce errors in the comprehensive distribution network analysis, lacking empirical fidelity. Without explicitly considering SDNets in OPF, the SDNet operating status cannot be accurately estimated. For instance, those works cannot detect and analyze over- and under-voltage problems in SDNets. Moreover, with the increasing proliferation of grid-edge resources, managing the voltage quality in SDNets is more challenging. 

Therefore, considering SDNet power flow in OPF is extremely vital for performing trustworthy, reliable, and practical distribution network analysis. There have been several works considering SDNets in distribution networks.  The works \cite{Transformer,DistributionSystem} demonstrate service transformer and triplex service line modeling in SDNets. The article \cite{Review}  surveys various technical requirements for integration of the roof-top PV into the existing low voltage distribution network. In \cite{rooftop}, researchers propose a hierarchical multilevel optimal power flow to minimize power losses in integrated primary-secondary distribution networks. But it does not provide a comprehensive treatment of the power flow in SDNets with service transformers. The work \cite{Qianzhi} proposes distributed optimal conservation voltage reduction for integrated primary-secondary distribution networks. However, this work simply regards a SDNet as a prosumer without considering service line segment and service transformer modeling, as well as power flow constraints. Service transformer and triplex service line power flow constraints in SDNets are non-linear and non-convex, resulting in challenges and difficulties in OPF solutions. The problem of incorporating service transformers and triplex service lines in SDNets into generalized OPF problems remains unresolved. 

To meet this gap, we propose an integrated primary-secondary distribution network OPF model, where service transformers and triplex service lines  at the SDNet are carefully considered. This OPF model enables considering distributed energy resources at both the primary and secondary distribution networks, thus bridging the gap between the PDNet optimization and SDNet optimization with increasing empirical fidelity. Compared with existing studies, the main contributions of this study are as follows:
\begin{itemize}
    \item A generalized integrated primary-secondary distribution network OPF is proposed. This OPF model carefully takes into account service transformer and  triplex service lines, leading to \textit{a holistic power system decision-making solution} at the distribution system level. 
    \item To resolve the nonlinear power flow constraints for service transformers, SOCP relaxation and linearization are applied to make service transformer constraints convex, respectively.
    \item Linearized power flow model is proposed for triplex service lines in the SDNet. In addition, its compact matrix-vector form is further developed by utilizing the graph structure of SDNet. Numerical studies show it provides a good estimate of real SDNet power flows by comparing its results with the nonlinear SDNet power flow.
\end{itemize}

\section{{Primary Distribution Network Model}}

\begin{figure*}
    \centering
    \includegraphics[width=7.4in]{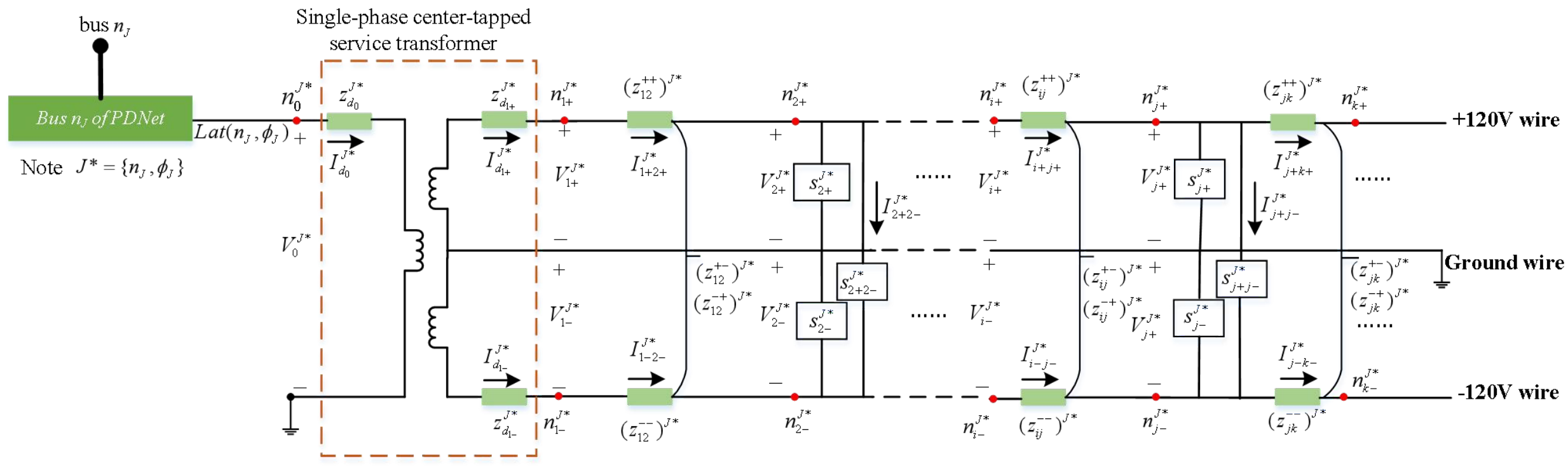}
    \caption{{The secondary distribution network $SDNet(n_J,\phi_J)$, connected to  bus $n_J$ of PDNet by a phase-$\phi_{J}$ lateral line $Lat(n_J,\phi_{J})$.}}
    \label{fig:sdnet}
\end{figure*}



The PDNet is an unbalanced radial primary distribution network PDNet. Let $\{n_0\}\bigcup\mathcal{N}$ denote the bus index set, where 0 is the index for the head bus and ${\mathcal{N}} = \{n_1,n_2,...,n_N\}$ is the index set for all non-head buses. For each bus $n_I\in\mathcal{N}$, let $\Phi_{I}\subseteq{\{a,b,c\}}$ denote the phase set of bus $n_I$. Let $v_{I}^{\phi_{I}}$, $p_{I}^{\phi_{I}}$ and $q_{I}^{\phi_{I}}$ denote the
phase-$\phi_{I}$ squared voltage magnitude, real and reactive power loads at bus $n_I$ in the PDNet, where $\phi_{I}\in\Phi_{I}$. And let $\bm{v}_I=[v_I^{\phi_I}]_{\phi_I\in\Phi_I}$, $\bm{p}_I=[p_I^{\phi_I}]_{\phi_I\in\Phi_I}$, $\bm{q}_I=[q_I^{\phi_I}]_{\phi_I\in\Phi_I}$. Let $\mathcal{L}$ denote the PDNet edge set. For each line segment $(n_I,n_J)\in\mathcal{L}$,
let $\bm{P}_{IJ}$, $\bm{Q}_{IJ}$ denote the real and reactive power flow vectors over the line segment $(n_I, n_J)$ in the PDNet, and let $\bm{Z}_{IJ}=\bm{R}_{IJ}+\text{j}\bm{X}_{IJ}$ denote the impedance matrix of line segment $(n_I, n_J)$  in the PDNet, including the self-impedance and mutual impedances of phases. For any line segment $(n_I, n_J)$ in the PDNet, the linearized PDNet power flow  \cite{LG} can be expressed as follows:

\begin{subequations}\label{eq:LPDNet}
    \begin{align}
    \bm{P}_{IJ}&=\sum_{(n_J,n_K)\in\mathcal{L}}\bm{P}_{JK}+\bm{p}_J\\
    \bm{Q}_{IJ}&=\sum_{(n_J,n_K)\in\mathcal{L}}\bm{Q}_{JK}+\bm{q}_J      \\
    \bm{v}_I&=\bm{v}_J+2\big[ \bm{\bar{R}}_{IJ}\bm{P}_{IJ}+\bm{\bar{X}}_{IJ}\bm{Q}_{IJ} \big] 
    \end{align}
\end{subequations}
where
\begin{align*}
    \bm{a} = ~[1, &e^{\rm-j2\pi/3}, e^{\rm j2\pi/3}]^T,~\bm{a}^{H} = \mbox{conjugate transpose of}~\bm{a}   \nonumber   \\
\bm{\bar{R}}_{ij} = & \text{real}(\bm{a}\bm{a}^{H})\odot{\bm{R}_{ij}}+\text{imag}(\bm{a}\bm{a}^{H})\odot{\bm{X}_{ij}}   \nonumber   \\
\bm{\bar{X}}_{ij} = & \text{real}(\bm{a}\bm{a}^{H})\odot{\bm{X}_{ij}}-\text{imag}(\bm{a}\bm{a}^{H})\odot{\bm{R}_{ij}}  \nonumber  \\
\odot = & ~\mbox{element-wise multiplication operator} \nonumber
\end{align*}

\section{{Secondary Distribution Network Model}}
\subsection{Overview}
This section  mainly introduces the SDNet model, consisting of {center-tapped} service transformers and  triplex service lines, and its nonlinear power flow.

\subsection{{Radial Secondary Distribution Network}}
{Hereafter, for ease of notation, we set $J*=\{n_J,\phi_{J}\}$, for any variable $(\cdot)^{J*}$ with superscript $J*$, this superscript $J*$ indicates the SDNet, $(\cdot)^{J*}$ belongs to, is $SDNet(n_J,\phi_{J})$.}  With respect to $SDNet(n_J,\phi_{J})$, the head-node $n_0^{J*}$ of $SDNet(n_J,\phi_{J})$ is electrically connected to bus $n_J$ of PDNet by a phase-$\phi_{J}$ lateral line $Lat(n_J,\phi_{J})$ and that includes a single-phase-$\phi_{J}$  center-tapped service transformer. An integrated primary-secondary distribution network illustration is shown in Fig.\ref{fig:TPSNet}. 

The center-tapped service transformer consists of three windings. As shown in Fig.\ref{fig:sdnet}, let ${z}_{d_{0}}^{J*}$, ${z}_{d_{1{+}}}^{J*}$, ${z}_{d_{1{-}}}^{J*}$ represent the individual winding impedances connected to the PDNet, +120V wire and -120V wire along the single-phase SDNet, respectively.  Let $I_{d_{0}}^{J*}$, $I_{d_{1+}}^{J*}$, $I_{d_{1-}}^{J*}$ denote the currents flowing through ${z}_{d_{0}}^{J*}$, ${z}_{d_{1{+}}}^{J*}$, ${z}_{d_{1{-}}}^{J*}$, 
$S_{d_{0}}^{J*}$, $S_{d_{1+}}^{J*}$, $S_{d_{1-}}^{J*}$ denote the power flows flowing through ${z}_{d_{0}}^{J*}$, ${z}_{d_{1{+}}}^{J*}$, ${z}_{d_{1{-}}}^{J*}$, respectively. We further define the current vector  $\bm{I}_{d_1}^{J*}\in\mathbb{C}^2$ and power flow vector  $\bm{S}_{d_1}^{J*}\in\mathbb{C}^2$, and  the impedance matrix $\bm{Z}_d^{J*}\in\mathbb{C}^{2\times 2}$  for the center-tapped service transformer as follows:
\begin{align*}
    \bm{I}_{d_1}^{J*}&=
    \begin{bmatrix}
    {I}_{d_{1{+}}}^{J*}\\
    {I}_{d_{1{-}}}^{J*}
    \end{bmatrix},
    \bm{S}_{d_1}^{J*}=\bm{P}_{d_1}^{J*}+\text{j}\bm{Q}_{d_1}^{J*}
    =\begin{bmatrix}
        S_{d_{1+}}^{J*}\\
        S_{d_{1-}}^{J*}
    \end{bmatrix}\\
    \bm{Z}_d^{J*}&=\bm{R}_d^{J*}+\text{j}\bm{X}_d^{J*}=
    \begin{bmatrix}
    {z}_{d_{1{+}}}^{J*}+{z}_{d_{0}}^{J*}&
    -{z}_{d_{0}}^{J*}\\
    {z}_{d_{0}}^{J*}&
    -({z}_{d_{1{-}}}^{J*}+{z}_{d_{0}}^{J*})
    \end{bmatrix}
\end{align*}

Let $\mathcal{N}^{J*}$ denote the index set of non-head nodes in $SDNet(n_J,\phi_{J})$, where $\mathcal{N}^{J*}=\{n_1^{J*},...,n_{N^{J*}}^{J*}\}$, $N^{J*}$ is the number of non-head nodes in $SDNet(n_J,\phi_{J})$. Note that $n_1^{J*}$ is the low-voltage side of service transformer, which is usually connected to end users by triplex service lines. {Each node $n^{J*}_i\in\mathcal{N}^{J*}\backslash\{n_1^{J*}\}$ is electrically connected to a collection of customers $C(n_i^{J*})$ by  a phase-$\phi_{J}$ triplex service line that includes three wires, i.e., +120V wire, the ground wire, -120V wire. Note that, as shown in Fig.\ref{fig:sdnet}, any non-head node $n^{J*}_i$ of $SDNet(n_J,\phi_{J})$ consists of 2-wire points, i.e., +120V-wire point $n^{J*}_{i+}$  and -120V-wire point $n^{J*}_{i-}$.} By definition of a radial network, each node located on a radial network can have at most one node that is an immediate predecessor. For each node $n_j^{J*}\in\mathcal{N}^{J*}\backslash\{n_1^{J*}\}$, let $np(n_j^{J*})\in\mathcal{N}^{J*}$ denote the node that immediately precedes node $n_j^{J*}$ along the radial SDNet. Let $\mathcal{L}^{J*}=\{(n_i^{J*},n_j^{J*})|n_i^{J*}=np(n_j^{J*}),n_j^{J*}\in\mathcal{N}^{J*}\backslash\{n_1^{J*}\}\}$ denote the edge set of the radial network $SDNet(n_J,\phi_{J})$,  i.e., the collection of all triplex service lines for $SDNet(n_J,\phi_{J})$. 



For the head node $n_0^{J*}$, let $V_{0}^{J*}$ denote the voltage at node $n_0^{J*}$, and $v_{{0}}^{J*}$ denote the squared voltage magnitude  of $V_{0}^{J*}$. And let $\bm{V}_0^{J*}\in\mathbb{C}^2$ denote the voltage vector for node $n_0^{J*}$:
\begin{align*}
    \bm{V}_{0}^{J*}=
    \begin{bmatrix}
    V_{0}^{J*}\\
    V_{0}^{J*}
    \end{bmatrix}
\end{align*}
As depicted in Fig.\ref{fig:sdnet}, for  non-head node $n_i^{J*}\in\mathcal{N}^{J*}$, let  $V_{{i+}}^{J*}$ and 
$V_{{i-}}^{J*}$ denote {the voltage difference} between its +120-wire point $n_{i+}^{J*}$ and the ground wire and {the voltage difference} between the ground wire and its -120-wire point $n_{i-}^{J*}$,  $v_{{i+}}^{J*}$ and 
$v_{{i-}}^{J*}$ denote the squared voltage magnitude values of $V_{{i+}}^{J*}$ and 
$V_{{i-}}^{J*}$. Let $\bm{V}_{i}^{J*}\in\mathbb{C}^2$ denote the voltage vector for node $n_i^{J*}\in\mathcal{N}^{J*}$:
\begin{align*}
    \bm{V}_{i}^{J*}=
     \begin{bmatrix}
    {V}_{i_+}^{J*}\\
    {V}_{i_-}^{J*}
    \end{bmatrix}, \forall{n_i^{J*}}\in\mathcal{N}^{J*}
\end{align*}
For node $n_i^{J*}\in\mathcal{N}^{J*}\backslash\{n_1^{J*}\}$,
let $s_{{i+}}^{J*}=p_{{i+}}^{J*}+\text{j}q_{{i+}}^{J*}$, $s_{{i-}}^{J*}=p_{{i-}}^{J*}+\text{j}q_{{i-}}^{J*}$, $s_{{i+i-}}^{J*}=p_{{i+i-}}^{J*}+\text{j}q_{{i+i-}}^{J*}$ denote the 120V net load connected to +120V-wire point $n_{i+}^{J*}$ and the ground wire, the 120V net load connected to the ground wire and -120V-wire point $n_{i-}^{J*}$, the 240V load connected to $n_{i+}^{J*}$ and $n_{i-}^{J*}$, respectively. As depicted in Fig.\ref{fig:sdnet}, let $I_{{i+i-}}^{J*}$ denote the current flowing from  $n_{i+}^{J*}$ to $n_{i-}^{J*}$ through the 240V load $s_{i+i-}^{J*}$.

For each  triplex service line segment $(n_i^{J*},n_j^{J*})\in\mathcal{L}^{J*}$, let $I_{{i+j+}}^{J*}$, $S_{i+j+}^{J*}=P_{{i+j+}}^{J*}+\text{j}Q_{i+j+}^{J*}$ denote the current and power flow from +120V-wire point $n_{i+}^{J*}$ to $n_{j+}^{J*}$; let $I_{i_-j_-}^{J*}$, $S_{i_-j_-}^{J*}=P_{i_-j_-}^{J*}+\text{j}Q_{i_-j_-}^{J*}$ denote the current and power flow from \text{-120V-wire} point $n_{i-}^{J*}$ to $n_{j-}^{J*}$. 
Next, let $\bm{I}_{ij}^{J*}\in\mathbb{C}^2$ and $\bm{S}_{ij}^{J*}\in\mathbb{C}^2$ denote its current and power flow vectors as follows:
\begin{align*}
    \bm{I}_{ij}^{J*}&=
     \begin{bmatrix}
    {I}_{i_+j_+}^{J*}\\
    {I}_{i_-j_-}^{J*}
    \end{bmatrix},
    \forall{(n_i^{J*},n_j^{J*})}\in\mathcal{L}^{J*},\\
    \bm{S}_{ij}^{J*}&=\bm{P}_{ij}^{J*}+\text{j}\bm{Q}_{ij}^{J*}=
     \begin{bmatrix}
    {S}_{i_+j_+}^{J*}\\
    {S}_{i_-j_-}^{J*}
    \end{bmatrix},
    \forall{(n_i^{J*},n_j^{J*})}\in\mathcal{L}^{J*}
\end{align*}
Finally, for ${(n_i^{J*},n_j^{J*})}\in\mathcal{L}^{J*} $, let $\bm{Z}_{ij}^{J*}\in\mathbb{C}^{2\times2}$ denote its impedance matrix of triplex service line segment:
\begin{equation*}
\begin{split}
   \bm{Z}_{ij}^{J*}=\bm{R}_{ij}^{J*}+\text{j}\bm{X}_{ij}^{J*}=
   \begin{bmatrix}
   ({z_{{ij}}^{++}})^{J*},(z_{{ij}}^{+-})^{J*}\\
   -({z_{{ij}}^{-+}})^{J*},-(z_{{ij}}^{--})^{J*}
   \end{bmatrix}
\end{split}
\end{equation*}
where $({z_{{ij}}^{++}})^{J*},(z_{{ij}}^{--})^{J*}$  denote the self-impedances of +120V and -120V wires, and $({z_{{ij}}^{+-}})^{J*},(z_{{ij}}^{-+})^{J*}$  denote the mutual impedances between +120V and -120V wires.
\subsection{Nonlinear SDNet Power Flow}
{According to \cite{Transformer,DistributionSystem}, service transformer power flow constraints are as follows:}
\begin{subequations}\label{eq:DT}
\begin{align}
    \bm{V}_{1}^{J*}&=\bm{V}_{0}^{J*}-\bm{Z}_d^{J*}\bm{I}_{d_1}^{J*}\\
     I_{d_{0}}^{J*}&=I_{d_{1+}}^{J*}-I_{d_{1-}}^{J*}
\end{align}
\end{subequations}
Substituting $I_{d_{0}}^{J*}=I_{d_{1+}}^{J*}-I_{d_{1-}}^{J*}$ into $S_{d_0}^{J*}=V_0^{J*}\overline{{I}_{d_0}^{J*}}$, the power flow $S_{d_0}^{J*}$ flowing through  $z_{d_0}^{J*}$ can be represented as follows:
\begin{equation}\label{eq:S0}
    S_{d_0}^{J*}=P_{d_0}^{J*}+\text{j}Q_{d_0}^{J*}=V_0^{J*}[\overline{I_{d_{1+}}^{J*}}-\overline{I_{d_{1-}}^{J*}}]
\end{equation}
According to \cite{Transformer,DistributionSystem}, the Ohm's law for triplex service line segment $(n_i^{J*},n_j^{J*})\in\mathcal{L}^{J*}$ is expressed as follows:
\begin{equation}\label{eq:OhmForLine}
    \bm{V}_{j}^{J*}=
    \bm{V}_{i}^{J*}
    -\bm{Z}_{ij}^{J*}\bm{I}_{ij}^{J*},
    \forall{(n_i^{J*},n_j^{J*})}\in\mathcal{L}^{J*}
\end{equation}
For any 240V-load $s_{i+i-}^{J*}$, it satisfies that: 
\begin{equation}\label{eq:240Vload}
    s_{{i+i-}}^{J*}=[V_{{i+}}^{J*}+V_{{i-}}^{J*}]\overline{I_{{i+i-}}^{J*}}, \forall{n_i^{J*}}\in\mathcal{N}^{J*}\backslash\{n_1^{J*}\}
\end{equation}
\section{Branch Flow Model for the SDNet with Service Transformers}
It is obvious that the nonlinear SDNet power flow model is non-convex, not suitable for performing OPF. For later convex optimization purposes, we first develop a branch flow model for the SDNet with service transformers, based on the nonlinear SDNet power flow model.
Define slack variables  $\bm{\ell}_{d}^{J*}\in\mathbb{C}^{2\times2}$, $\bm{W}_d^{J*}\in\mathbb{C}^{2\times2}$ as follows:
\begin{align*}
    \bm{\ell}_{d}^{J*}&=\bm{I}_{d_1}^{J*}[\bm{I}_{d_1}^{J*}]^H=
    \begin{bmatrix}
        \ell_{d_{1+}}^{J*},\ell_{d_{1+-}}^{J*}\\
        \overline{\ell_{d_{1+-}}^{J*}},\ell_{d_{1-}}^{J*}
    \end{bmatrix}\\
    \bm{W}_d^{J*}&=\bm{V}_0^{J*}[\bm{I}_{d_1}^{J*}]^H=
    \begin{bmatrix}
        S_{d_{0+}}^{J*},S_{d_{0-}}^{J*}\\
        S_{d_{0+}}^{J*},S_{d_{0-}}^{J*}
    \end{bmatrix}
\end{align*}
where:
\begin{subequations}\label{eq:ldSd}
    \begin{align}
        {\ell}_{d_{1+}}^{J*}&={I}_{d_{1+}}^{J*}\overline{{I}_{d_{1+}}^{J*}}\\
         {\ell}_{d_{1-}}^{J*}&={I}_{d_{1-}}^{J*}\overline{{I}_{d_{1-}}^{J*}}
        \\{\ell}_{d_{1+-}}^{J*}&={I}_{d_{1+}}^{J*}\overline{{I}_{d_{1-}}^{J*}}\\
        S_{d_{0+}}^{J*}&=P_{d_{0+}}^{J*}+\text{j}Q_{d_{0+}}^{J*}=V_{0}^{J*}\overline{I_{d_{1+}}^{J*}}\\
        S_{d_{0-}}^{J*}&=P_{d_{0-}}^{J*}+\text{j}Q_{d_{0-}}^{J*}=V_{0}^{J*}\overline{I_{d_{1-}}^{J*}}
    \end{align}
\end{subequations}
We further introduce $\bm{U}_{0}^{J*}\in\mathbb{R}^{2\times2}$, and $\bm{v}_{0}^{J*}\in\mathbb{R}^{2}$ as follows: 
\begin{align*}
    \bm{U}_{0}^{J*}&=\bm{V}_{0}^{J*}[\bm{V}_{0}^{J*}]^H=
        \begin{bmatrix}
    {v}_0^{J*},
    &{v}_0^{J*}
    \\
    {v}_0^{J*},
    &{v}_0^{J*}
    \end{bmatrix}\\
    \bm{v}_0^{J*}&=
    \text{diag}(\bm{U}_{0}^{J*})=
    \begin{bmatrix}
        v_0^{J*}\\
        v_0^{J*}
    \end{bmatrix}
\end{align*}
For any triplex service line segment $\forall (n_i^{J*},n_j^{J*})\in\mathcal{L}^{J*}$, we introduce slack variables $\bm{\ell}_{ij}^{J*}\in\mathbb{C}^{2\times2}$, $\bm{W}_{ij}^{J*}\in\mathbb{C}^{2\times2}$:
\begin{subequations}\label{eq:lWij}\footnotesize
\begin{align}
      \nonumber\bm{\ell}_{ij}^{J*}&=\bm{I}_{ij}^{J*}[\bm{I}_{ij}^{J*}]^H\\
    &=
    \begin{bmatrix}
    {I}_{i+j+}^{J*}\overline{{I}_{i+j+}^{J*}}&{I}_{i+j+}^{J*}\overline{{I}_{i-j-}^{J*}}\\
    {I}_{i-j-}^{J*}\overline{{I}_{i+j+}^{J*}}&{I}_{i-j-}^{J*}\overline{{I}_{i-j-}^{J*}}
    \end{bmatrix}
    \\
    \nonumber&\bm{W}_{ij}^{J*}=\bm{V}_{i}^{J*}[\bm{I}_{ij}^{J*}]^H\\
    &=
    \begin{bmatrix}
    {V}_{i+}^{J*}\overline{{I}_{i+j+}^{J*}}&{V}_{i+}^{J*}\overline{{I}_{i-j-}^{J*}}\\
    {V}_{i-}^{J*}\overline{{I}_{i+j+}^{J*}}&{V}_{i-}^{J*}\overline{{I}_{i-j-}^{J*}}
    \end{bmatrix}
    =
    \begin{bmatrix}
    S_{i+j+}^{J*}&{V}_{i+}^{J*}\overline{{I}_{i-j-}^{J*}}\\
    {V}_{i-}^{J*}\overline{{I}_{i+j+}^{J*}}&
    S_{i-j-}^{J*}
    \end{bmatrix}
\end{align}
\end{subequations}
We also introduce $\bm{U}_i^{J*}\in\mathbb{C}^{2\times2}$, and $\bm{v}_i^{J*}\in\mathbb{R}^{2\times1}$ for $\forall (n_i^{J*},n_j^{J*})\in\mathcal{L}^{J*}$:
\begin{align*}
     \bm{U}_i^{J*}&=\bm{V}_i^{J*}[\bm{V}_i^{J*}]^H=
    \begin{bmatrix}
    {V}_{i+}^{J*}\overline{{V}_{i+}^{J*}}&{V}_{i+}^{J*}\overline{{V}_{i-}^{J*}}
    \\
    {V}_{i-}^{J*}\overline{{V}_{i+}^{J*}}
    &{V}_{i-}^{J*}\overline{{V}_{i-}^{J*}}
    \end{bmatrix}\\
    \bm{v}_i^{J*}&=\text{diag}( \bm{U}_i^{J*})
    =
    \begin{bmatrix}
        v_{i+}^{J*}\\
        v_{i-}^{J*}
    \end{bmatrix}
\end{align*}
With respect to (\ref{eq:DT}a), multiply both sides by their Hermitian transposes, 
i.e., $[\bm{V}_{1}^{J*}]^H, [\bm{V}_{0}^{J*}-\bm{Z}_d^{J*}\bm{I}_{d}^{J*}]^H$, we have:
\begin{equation}\label{eq:v1}
\begin{split}
\bm{U}_{1}^{J*}=\bm{U}_{0}^{J*}-\bm{W}_d^{J*}[\bm{Z}_d^{J*}]^H-\bm{Z}_d^{J*}[\bm{W}_d^{J*}]^H+\bm{Z}_d^{J*}\bm{\ell}_{d}^{J*}[\bm{Z}_d^{J*}]^H
\end{split}
\end{equation}
Apply $\text{diag}(\cdot)$ to (\ref{eq:v1}), the squared voltage magnitude relationship for the service transformer is:
\begin{equation}\label{eq:nonlinearv01}
\begin{split}
    \bm{v}_{1}^{J*}&=\bm{v}_{0}^{J*}-\text{diag}(\bm{W}_d^{J*}[\bm{Z}_d^{J*}]^H)-\text{diag}(\bm{Z}_d^{J*}[\bm{W}_d^{J*}]^H)\\
    &+\text{diag}(\bm{Z}_d^{J*}\bm{\ell}_{d}^{J*}[\bm{Z}_d^{J*}]^H)
\end{split}
\end{equation}
And (\ref{eq:S0}) can be equivalently denoted by:
\begin{equation}\label{eq:NewS0}
    {S}_{d_0}^{J*}
    =
    \begin{bmatrix}
    1, -1
    \end{bmatrix}
    \text{diag}(\bm{W}_d^{J*})
    =S_{d_{0+}}^{J*}-S_{d_{0-}}^{J*}
\end{equation}
With respect to (\ref{eq:OhmForLine}), multiply both sides by their Hermitian transposes, i.e.,  $[\bm{V}_j^{J*}]^H, [\bm{V}_{i}^{J*}-\bm{Z}_{ij}^{J*}\bm{I}_{ij}^{J*}]^H$, then we have:
\begin{equation}\label{eq:vk}
\begin{split}
    \bm{U}_{j}^{J*}=&[\bm{V}_{i}^{J*}
    -\bm{Z}_{ij}^{J*}\bm{I}_{ij}^{J*}][\bm{V}_{i}^{J*}
    -\bm{Z}_{ij}^{J*}\bm{I}_{ij}^{J*}]^H\\
    =&\bm{U}_{i}^{J*}-\bm{W}_{ij}^{J*}[\bm{Z}_{ij}^{J*}]^H-\bm{Z}_{ij}^{J*}[\bm{W}_{ij}^{J*}]^H\\&+\bm{Z}_{ij}^{J*}\bm{\ell}_{ij}^{J*}[\bm{Z}_{ij}^{J*}]^H, \forall{(n_i^{J*},n_j^{J*})}\in\mathcal{L}^{J*}
\end{split}
\end{equation}
Apply $\text{diag}(\cdot)$ to (\ref{eq:vk}), the squared voltage magnitude relationship for triplex service lines  is:
\begin{equation}\label{eq:nonlinearvivj}
\begin{split}
    \bm{v}_{j}^{J*}&=\bm{v}_{i}^{J*}-\text{diag}(\bm{W}_{ij}^{J*}[\bm{Z}_{ij}^{J*}]^H)-\text{diag}(\bm{Z}_{ij}^{J*}[\bm{W}_{ij}^{J*}]^H)\\&+\text{diag}(\bm{Z}_{ij}^{J*}\bm{\ell}_{ij}^{J*}[\bm{Z}_{ij}^{J*}]^H), \forall{(n_i^{J*},n_j^{J*})}\in\mathcal{L}^{J*}
\end{split}
\end{equation}
{Multiplying by the Hermitian transpose \footnote{{Hermitian transposes, indicated by superscript $H$, take the matrix transpose and then take the complex conjugate of each entry in the matrix.}
} of $\bm{I}_{ij}^{J*}$, i.e., $[\bm{I}_{ij}^{J*}]^{H}$, on both sides of (\ref{eq:OhmForLine})}, then we have:
\begin{equation}\label{eq:VI}
\begin{split}
    \bm{V}_{j}^{J*}[\bm{I}_{ij}^{J*}]^{H}&=
    \bm{V}_{i}^{J*}[\bm{I}_{ij}^{J*}]^{H}
    -\bm{Z}_{ij}^{J*}\bm{I}_{ij}^{J*}[\bm{I}_{ij}^{J*}]^{H}\\
    &=\bm{W}_{ij}^{J*}-\bm{Z}_{ij}^{J*}\bm{\ell}_{ij}^{J*}, \forall{(n_i^{J*},n_j^{J*})}\in\mathcal{L}^{J*}
\end{split}
\end{equation}
Note that $\text{diag}(\bm{V}_{j}^{J*}[\bm{I}_{ij}^{J*}]^{H})$ is the receiving-end power flow on triplex service line segment $(n_i^{J*},n_j^{J*})\in\mathcal{L}^{J*}$. Thus,
the power balance for triplex service lines is
expressed as follows:
\begin{equation}
\begin{split}
    &\text{diag}(\bm{W}_{ij}^{J*}-\bm{Z}_{ij}^{J*}\bm{\ell}_{ij}^{J*})=\sum_{(n_j^{J*},n_k^{J*})\in\mathcal{L}^{J*}}\text{diag}([\bm{W}_{jk}^{J*}]^{H})\\&+
    \begin{bmatrix}
    s_{j+}^{J*}\\
    s_{j-}^{J*}
    \end{bmatrix}
    +
    \begin{bmatrix}
    V_{j+}^{J*}\overline{I_{j+j-}^{J*}}\\
    -V_{j-}^{J*}\overline{I_{j+j-}^{J*}}
    \end{bmatrix},\forall n_j^{J*}\in\mathcal{N}^{J*}\backslash\{n_1^{J*}\}
\end{split}
\end{equation}
That is:
\begin{equation}\label{eq:PBforSDNet}
\begin{split}
&\bm{S}_{ij}^{J*}-\text{diag}(\bm{Z}_{ij}^{J*}\bm{\ell}_{ij}^{J*})=\sum_{(n_j^{J*},n_k^{J*})\in\mathcal{L}^{J*}}\bm{S}_{jk}^{J*}+
    \begin{bmatrix}
    s_{j+}^{J*}\\
    s_{j-}^{J*}
    \end{bmatrix}
    \\&+
    \begin{bmatrix}
    V_{j+}^{J*}\overline{I_{j+j-}^{J*}}\\
    -V_{j-}^{J*}\overline{I_{j+j-}^{J*}}
    \end{bmatrix},\forall n_j^{J*}\in\mathcal{N}^{J*}\backslash\{n_1^{J*}\}
\end{split}
\end{equation}
The power flow relationship between the service transformer and triplex service lines is:
\begin{equation}\label{eq:relationship}
    \bm{S}_{d_1}^{J*}=\sum_{(n_1^{J*},n_j^{J*})\in\mathcal{L}^{J*}}\bm{S}_{1j}^{J*}
\end{equation}
In short, SDNet branch flow  constraints include service transformer constraints and triplex service line constraints. Service transformer constraints include:
\begin{itemize}
    \item Slack variable constraints: (\ref{eq:ldSd})
    \item Voltage and power balance constraints for the single-phase center-tapped service transformer: (\ref{eq:nonlinearv01}),(\ref{eq:NewS0})
\end{itemize}
Triplex service line constraints include:
\begin{itemize}
    \item The 240V-load constraints: (\ref{eq:240Vload})
    \item Slack variable constraints: (\ref{eq:lWij})
    \item The voltage power balance constraints for triplex service lines: (\ref{eq:nonlinearvivj}), (\ref{eq:PBforSDNet})
    \item The power balance between the single-phase center-tapped service transformer and triplex service lines: (\ref{eq:relationship})
\end{itemize}


\section{Convex Relaxation and Linearization for Service Transformer Constraints}
This section considers convex relaxation and linearization for service transformer constraints.
\subsection{SOCP Relaxation for Service Transformers}
The nonlinear equation (\ref{eq:ldSd}) can be expressed as follows:
\begin{subequations}\label{eq:ldSd+Sd-}
\begin{align}
    \text{real}(\ell_{d_{1+-}}^{J*})^2+\text{imag}(\ell_{d_{1+-}}^{J*})^2={\ell}_{d_{1+}}^{J*}{\ell}_{d_{1-}}^{J*}\\
    \text{real}(S_{d_{0+}}^{J*})^2+\text{imag}(S_{d_{0+}}^{J*})^2=v_0^{J*}{\ell}_{d_{1+}}^{J*}\\
    \text{real}(S_{d_{0-}}^{J*})^2+\text{imag}(S_{d_{0-}}^{J*})^2=v_0^{J*}{\ell}_{d_{1-}}^{J*}\\
     \text{imag}(\ell_{d_{1+}}^{J*})=0,\text{imag}(\ell_{d_{1-}}^{J*})=0
\end{align}
\end{subequations}
The nonlinear equation (\ref{eq:ldSd+Sd-}) is still non-convex, we apply the SOCP relaxation to (\ref{eq:ldSd+Sd-}) as follows:
\begin{subequations}\label{eq:SOCP}
\begin{align}
    \text{real}(\ell_{d_{1+-}}^{J*})^2+\text{imag}(\ell_{d_{1+-}}^{J*})^2\leq{\ell}_{d_{1+}}^{J*}{\ell}_{d_{1-}}^{J*}\\
    \text{real}(S_{d_{0+}}^{J*})^2+\text{imag}(S_{d_{0+}}^{J*})^2\leq{v}_0^{J*}{\ell}_{d_{1+}}^{J*}\\
    \text{real}(S_{d_{0-}}^{J*})^2+\text{imag}(S_{d_{0-}}^{J*})^2\leq{v}_0^{J*}{\ell}_{d_{1-}}^{J*}\\
    \text{imag}(\ell_{d_{1+}}^{J*})=0,\text{imag}(\ell_{d_{1-}}^{J*})=0
\end{align}
\end{subequations}
Note that (\ref{eq:SOCP}) can be rewritten as the standard second-order cone, which is convex:
\begin{subequations}
\begin{align}
    \norm{
    \begin{bmatrix}
        2\text{real}(\ell_{d_{1+-}}^{J*})\\
        2\text{imag}(\ell_{d_{1+-}}^{J*})\\
        {\ell}_{d_{1+}}^{J*}-{\ell}_{d_{1-}}^{J*}
    \end{bmatrix}
    }_2\leq
    \norm{{\ell}_{d_{1+}}^{J*}+{\ell}_{d_{1-}}^{J*}}_2
\end{align}
\begin{align}
    \norm{
    \begin{bmatrix}
        2\text{real}(S_{d_{0+}}^{J*})\\
        2\text{imag}(S_{d_{0+}}^{J*})\\
        v_0^{J*}-{\ell}_{d_{1+}}^{J*}
    \end{bmatrix}
    }_2\leq
    \norm{v_0^{J*}+{\ell}_{d_{1+}}^{J*}}_2
\end{align}
\begin{align}
    \norm{
    \begin{bmatrix}
        2\text{real}(S_{d_{0-}}^{J*})\\
        2\text{imag}(S_{d_{0-}}^{J*})\\
        v_0^{J*}-{\ell}_{d_{1-}}^{J*}
    \end{bmatrix}
    }_2\leq
    \norm{v_0^{J*}+{\ell}_{d_{1-}}^{J*}}_2\\
    \text{imag}(\ell_{d_{1+}}^{J*})=0,\text{imag}(\ell_{d_{1-}}^{J*})=0
\end{align}
\end{subequations}
The SOCP-based service transformer constraints (SOCP-ST) are defined:
\begin{align*}
    \text{SOCP-ST}=\{(\ref{eq:nonlinearv01}),(\ref{eq:NewS0}),(\ref{eq:SOCP})\}
\end{align*}

\subsection{Linearization for Service Transformers}
This part illustrates the linearization for  non-convex service transformer constraints, which is based on [Assumption 1].

\textbf{[Assumption 1]}: The service transformer losses are relatively small compared to its power flows, i.e., $\bm{Z}_{d}^{J*}\bm{\ell}_{d}^{J*}<<\bm{W}_{d}^{J*}$.

Multiplying both sides of (\ref{eq:DT}) by  $(\bm{I}_{d_1}^{J*})^H$ and applying $\text{diag}(\cdot)$ to it, we have: 
\begin{equation}\label{eq:SMZ}
\begin{split}
    \text{diag}(\bm{V}_{1}^{J*}(\bm{I}_{d_1}^{J*})^H)&=\text{diag}(\bm{V}_{0}^{J*}(\bm{I}_{d_1}^{J*})^H)-\text{diag}(\bm{Z}_d^{J*}\bm{\ell}_{d}^{J*})\\
    \bm{S}_{d_1}^{J*}&=\text{diag}(\bm{W}_d^{J*})-\text{diag}(\bm{Z}_d^{J*}\bm{\ell}_{d}^{J*})
\end{split}
\end{equation}
Neglecting $\text{diag}(\bm{Z}_d^{J*}\bm{\ell}_{d}^{J*})$ in (\ref{eq:SMZ}) based on [Assumption 1], it follows that $\bm{S}_{d_1}^{J*}=\text{diag}(\bm{W}^{J*}_d)$, i.e., :
\begin{equation}\label{eq:linearSM}
     \bm{P}_{d_1}^{J*}=
     \begin{bmatrix}
         P_{d_{0+}}^{J*}\\
         P_{d_{0-}}^{J*}
     \end{bmatrix},
     \bm{Q}_{d_1}^{J*}=
     \begin{bmatrix}
         Q_{d_{0+}}^{J*}\\
         Q_{d_{0-}}^{J*}
     \end{bmatrix}
\end{equation}
From [Assumption 1], we  neglect $\text{diag}(\bm{Z}_{ij}^{J*}\bm{\ell}_{ij}^{J*}[\bm{Z}_{ij}^{J*}]^H)$ in (\ref{eq:nonlinearv01}), then it follows:
\begin{equation}\label{eq:v10}
\begin{split}
    &\bm{v}_{1}^{J*}=\bm{v}_{0}^{J*}-\text{diag}(\bm{W}_d^{J*}[\bm{Z}_d^{J*}]^H)-\text{diag}(\bm{Z}_d^{J*}[\bm{W}_d^{J*}]^H)\\
    &=\bm{v}_{0}^{J*}-2\bm{R}_{d}^{J*}
    \text{real}(\text{diag}(\bm{W}^{J*}_d))
    -2\bm{X}_d^{J*}\text{imag}(\text{diag}(\bm{W}^{J*}_d))\\
    &=\bm{v}_{0}^{J*}-2\bm{R}_{d}^{J*}
    \begin{bmatrix}
        P_{d_{0+}}^{J*}\\
        P_{d_{0-}}^{J*}
    \end{bmatrix}
    -2\bm{X}_{d}^{J*}
    \begin{bmatrix}
        Q_{d_{0+}}^{J*}\\
        Q_{d_{0-}}^{J*}
    \end{bmatrix}
\end{split}
\end{equation}
Substituting (\ref{eq:linearSM}) into (\ref{eq:v10}), we have:
\begin{equation}\label{eq:LST}
\bm{v}_{1}^{J*}=\bm{v}_{0}^{J*}-2\bm{R}_{d}^{J*}\bm{P}_{d_1}^{J*}-2\bm{X}_d^{J*}\bm{Q}_{d_1}^{J*}
\end{equation}
The linearized service transformer constraints (L-ST) are defined:
\begin{align*}
    \text{L-ST}=\{(\ref{eq:linearSM}),(\ref{eq:LST})\}
\end{align*}


\section{Linearization for Triplex Service Lines}
\label{sec:LTS}
\subsection{Linearized Triplex Service Line Power Flow}
The linearized triplex service line power flow is proposed, based on the following \textbf{[Assumption 2]}:
\begin{itemize}
    \item The voltage unbalances on each SDNet node $\forall{n_{i}^{J*}}\in\mathcal{N}^{J*}$ are not severe: $V_{i+}^{J*}\approx{V}_{i-}^{J*}$. 
    \item The line losses are relatively small compared to the power flows, i.e., $\bm{Z}_{ij}^{J*}\bm{\ell}_{ij}^{J*}<<\bm{M}_{ij}^{J*}$.
\end{itemize}
From [Assumption 2], it indicates from (\ref{eq:240Vload}) and (\ref{eq:lWij}) that:
\begin{subequations}\small\label{eq:A1}
\begin{align}
V_{i+}^{J*}\overline{I_{i+i-}^{J*}}&={V}_{i-}^{J*}\overline{I_{i+i-}^{J*}}=\frac{s_{i+i-}^{J*}}{2},\forall{n_i^{J*}}\in\mathcal{N}^{J*}\backslash\{n_1^{J*}\}\\ 
 \bm{W}_{ij}^{J*}&=
    \begin{bmatrix}
        S_{i+j+}^{J*}, S_{i-j-}^{J*}\\
        S_{i+j+}^{J*}, S_{i-j-}^{J*}
    \end{bmatrix},
    \forall (n_i^{J*},n_j^{J*})\in\mathcal{L}^{J*}
\end{align}
\end{subequations}
Neglecting $\text{diag}(\bm{Z}_{ij}^{J*}\bm{\ell}_{ij}^{J*}[\bm{Z}_{ij}^{J*}]^H)$ in (\ref{eq:nonlinearvivj}), and substituting (\ref{eq:A1}b) into (\ref{eq:nonlinearvivj}), then we have:
\begin{equation}\label{eq:linearizedv}
\begin{split}
    \bm{v}_{j}^{J*}
=\bm{v}_{i}^{J*}-2\bm{R}_{ij}^{J*}\bm{P}_{ij}^{J*}-2\bm{X}_{ij}^{J*}\bm{Q}_{ij}^{J*}, \forall{(n_i^{J*},n_j^{J*})}\in\mathcal{L}^{J*}
\end{split}
\end{equation}
Substituting (\ref{eq:A1}a) into (\ref{eq:PBforSDNet}) and neglecting $\text{diag}(\bm{Z}_{ij}^{J*}\ell_{ij}^{J*})$ in (\ref{eq:PBforSDNet}) due to [Assumption 2], we have:
\begin{equation}
\begin{split}
\bm{S}_{ij}^{J*}&=\sum_{(n_j^{J*},n_k^{J*})\in\mathcal{L}^{J*}}\bm{S}_{jk}^{J*}+
    \begin{bmatrix}
    s_{j+}^{J*}\\
    s_{j-}^{J*}
    \end{bmatrix}
    \\&+
    \begin{bmatrix}
    \frac{s_{j+j-}^{J*}}{2}\\
    -\frac{s_{j+j-}^{J*}}{2}
    \end{bmatrix},\forall n_j^{J*}\in\mathcal{N}^{J*}\backslash\{n_1^{J*}\}
\end{split}
\end{equation}
That is:
\begin{subequations}\label{eq:linearizedpower}
\begin{align}
\nonumber\bm{P}_{ij}^{J*}&=\sum_{(n_j^{J*},n_k^{J*})\in\mathcal{L}^{J*}}\bm{P}_{jk}^{J*}+
    \begin{bmatrix}
    p_{j+}^{J*}\\
    p_{j-}^{J*}
    \end{bmatrix}
    \\&+
    \begin{bmatrix}
    \frac{p_{j+j-}^{J*}}{2}\\
    -\frac{p_{j+j-}^{J*}}{2}
    \end{bmatrix},\forall n_j^{J*}\in\mathcal{N}^{J*}\backslash\{n_1^{J*}\}\\
\nonumber\bm{Q}_{ij}^{J*}&=\sum_{(n_j^{J*},n_k^{J*})\in\mathcal{L}^{J*}}\bm{Q}_{jk}^{J*}+
    \begin{bmatrix}
    q_{j+}^{J*}\\
    q_{j-}^{J*}
    \end{bmatrix}
    \\&+
    \begin{bmatrix}
    \frac{q_{j+j-}^{J*}}{2}\\
    -\frac{q_{j+j-}^{J*}}{2}
    \end{bmatrix},\forall n_j^{J*}\in\mathcal{N}^{J*}\backslash\{n_1^{J*}\}  
\end{align}
\end{subequations}
Besides, (\ref{eq:relationship}) can be equivalently written as follows:
\begin{subequations}\label{eq:PdQd}
    \begin{align}
    \bm{P}_{d_1}^{J*}&=\sum_{(n_1^{J*},n_j^{J*})\in\mathcal{L}^{J*}}\bm{P}_{1j}^{J*}\\
    \bm{Q}_{d_1}^{J*}&=\sum_{(n_1^{J*},n_j^{J*})\in\mathcal{L}^{J*}}\bm{Q}_{1j}^{J*}
    \end{align}
\end{subequations}
The linearized triplex service line constraints (L-TSL) are defined:
\begin{align*}
    \text{L-TSL}=\{(\ref{eq:linearizedv}),(\ref{eq:linearizedpower}),(\ref{eq:PdQd})\}
\end{align*}

\textbf{Case studies in Section \ref{sec:Case}-A show that the L-TSL provides a good estimate of power flows and voltages.} The L-TSL provides a linearized power flow model for SDNet.

\subsection{Graph-Based Compact Power Flow Representation}
Consider, first, the standard matrix representation $\bm{\bar{M}^{J*}}=[\bm{m}_1^{J*}, [\bm{M}^{J*}]^T]^T\in\mathbb{R}^{N^{J*}\times (N^{J*}-1)}$ for the incidence matrix of +120V-wire circuit in the SDNet. The rows of this matrix correspond to  +120V-wire points, and the columns of this matrix correspond to +120V line segments. The entries of the matrix indicate, for each +120V-wire point and line segment, whether or not the +120-wire point is a ``from'' point or ``to'' node for this +120V line segment.

{More precisely, $\bm{\bar{M}}^{J*}$ with an entry 1 for each ``from'' point and -1 for each ``to'' node takes the following form:}
\begin{equation}
    {\bm{\bar{M}}}^{J*} = \begin{bmatrix}
            \mathbb{J}\big(1,(np(n_2^{J*}),n_2^{J*})\big)&...&\mathbb{J}\big(1,(np(n_N^{J*}),n_N^{J*})\big)\\
            \mathbb{J}\big(2,(np(n_2^{J*}),n_2^{J*})\big)&...&\mathbb{J}\big(2,(np(n_N^{J*}),n_N^{J*})\big)\\
            \vdots&\vdots&\vdots\\
            \mathbb{J}\big(N^{J*},(np(n_2^{J*}),n_2^{J*})\big)&...&\mathbb{J}\big(N^{J*},(np(n_N^{J*}),n_N^{J*})\big)
            \end{bmatrix}
\end{equation}
where $\mathbb{J}(\cdot)$ is an indicator function defined as:
\begin{align*}
\mathbb{J}\big(i,(np(n_j^{J*}),n_j^{J*})\big)=
    \begin{cases}
     1 & \mbox{if}~n_i^{J*}=np(n_j^{J*})\\
    -1 & \mbox{if}~n_i^{J*}=n_j^{J*}\\
     0 & \mbox{otherwise}
    \end{cases}
\end{align*}
The first row $[\bm{m}_1^{J*}]^T$ of the matrix $\bm{\bar{M}}^{J*}$ represents the connection structure between +120V-wire point $n_{1+}^{J*}$ and the line segments in $\mathcal{L}^{J*}$; the remaining submatrix, denoted by $\bm{M}^{J*}$, represents the connection structure between the remaining +120V-wire points and +120V line segments. The tree topology of SDNet ensures that it is a connected graph, where $\bm{M}^{J*}$ is a square matrix with full rank \cite{Graph}, and thus invertible.  

Next, the incidence matrix $\bm{\bar{M}^{J*}}$ can be extended to the triplex line segments, consisting of +120V-wire circuit and -120V-wire circuit, in the SDNet. An extended incidence matrix \footnote{One simple example of incidence matrix construction for secondary distribution networks is given in Appendix A.} $\bm{\bar{A}}^{J*}=[\bm{A}_1^{J*}, [\bm{A}^{J*}]^T]^T\in\mathbb{R}^{(2*N^{J*})\times (2*N^{J*}-2)}$ for triplex service lines is constructed as follows:
 \begin{equation}
 \begin{split}
    &\bar{\bm A}^{J*}={\bm{\bar{M}}}^{J*}\otimes{\textbf{I}_{2}}
 \end{split}
\end{equation}
where 
\begin{align*}
              & \text{I}_{2} ~~=~ 2 \times 2 ~ \mbox{identity matrix};   \\
              & \otimes ~=~ \mbox{Kronecker product operation.}
\end{align*}
For $\forall n_i^{J*}\in\mathcal{N}^{J*}\backslash\{n_1^{J*}\}$, we define $\bm{p}_i^{J*}\in\mathbb{R}^2$ and $\bm{q}_i^{J*}\in\mathbb{R}^2$ as follows:
\begin{subequations}\label{eq:pqJ}
    \begin{align}
    \bm{p}_i^{J*}=
    \begin{bmatrix}
        p_{i+}^{J*}\\
        p_{i-}^{J*}
    \end{bmatrix}
    +
    \begin{bmatrix}
        \frac{p_{i+i-}^{J*}}{2}\\
        -\frac{p_{i+i-}^{J*}}{2}
    \end{bmatrix}\\
    \bm{q}_i^{J*}=
    \begin{bmatrix}
        q_{i+}^{J*}\\
        q_{i-}^{J*}
    \end{bmatrix}
    +
    \begin{bmatrix}
        \frac{q_{i+i-}^{J*}}{2}\\
        -\frac{q_{i+i-}^{J*}}{2}
    \end{bmatrix}
    \end{align}
\end{subequations}
Define the following vectors:
\begin{align*}
    \bm{v}^{J*}&=[~[\bm{v}_2^{J*}]^T,...,[\bm{v}_N^{J*}]^T~]^T\\
    \bm{p}^{J*}&=[~[\bm{p}_2^{J*}]^T,...,[\bm{p}_N^{J*}]^T~]^T\\
     \bm{q}^{J*}&=[~[\bm{q}_2^{J*}]^T,...,[\bm{q}_N^{J*}]^T~]^T\\
     \bm{P}^{J*}&=[~[\bm{P}_{np(2)2}^{J*}]^T, [\bm{P}_{np(3)3}^{J*}]^T,..., [\bm{P}_{np(N)N}^{J*}]^T~]^T\\
    \bm{Q}^{J*}&=[~[\bm{Q}_{np(2)2}^{J*}]^T, [\bm{Q}_{np(3)3}^{J*}]^T,..., [\bm{Q}_{np(N)N}^{J*}]^T~]^T
\end{align*}
where $\bm{P}_{np(j)j}^{J*},~\bm{Q}_{np(j)j}^{J*}$ denote the real and reactive power flow for the triplex service line $\ell_j^{J*}=(np(n_j^{J*}),n_j^{J*})$.

Resistances and reactances for triplex lines in $\mathcal{L}^{J*}$ are compactly denoted by $2(N^{J*}-1)\times 2(N^{J*}-1)$ block diagonal matrices $\bm{D}_r^{J*}$ and $\bm{D}_x^{J*}$ as:
\begin{align*}
    \bm{D}_r^{J*}&=\text{diag}(\bm{R}_{np(2)2}^{J*},\bm{R}_{np(3)3}^{J*},...,\bm{R}_{np(N)N}^{J*})\\
    \bm{D}_x^{J*}&=\text{diag}(\bm{X}_{np(2)2^{J*}},\bm{X}_{np(3)3}^{J*},...,\bm{X}_{np(N)N}^{J*})
\end{align*}
where $\bm{R}_{np(j)j}^{J*},~\bm{X}_{np(j)j}^{J*}$ denote the resistance and reactance matrix for the triplex service line $\ell_j^{J*}=(np(n_j^{J*}),n_j^{J*})$.

In this case, (\ref{eq:linearizedv}) and (\ref{eq:linearizedpower}) can be expressed as follows:
   \begin{subequations}\label{eq:compactPL}
                         \begin{align}
\bm{A}^{J*}\bm{P}^{J*}&=-\bm{p}^{J*}\\
\bm{A}^{J*}\bm{Q}^{J*}&=-\bm{q}^{J*}\\
\begin{bmatrix}{\bm{A}}_1^{J*}\, \,[\bm{A}^{J*}]^T\end{bmatrix}\begin{bmatrix}{\bm{v}_1^{J*}}\\{\bm{v}^{J*}}\end{bmatrix}&=2\Big(\bm{D}_r^{J*}\bm{P}^{J*}+\bm{D}_x^{J*}\bm{Q}^{J*}\Big)
                       \end{align}
                        \end{subequations}
The compact linearized triplex service line constraints (C-L-TSL) are defined:
\begin{align*}
    \text{C-L-TSL}=\{(\ref{eq:PdQd}),(\ref{eq:pqJ}),(\ref{eq:compactPL})\}
\end{align*}

\textbf{[Remark]:} It is worth mentioning that $\bm{A}^{J*}$ is a square matrix with full rank and invertible since  it satisfies $\bm{A}^{J*}=\bm{M}^{J*}\otimes{\textbf{I}_{2}}$, where $\bm{M}^{J*}$ is a square matrix with full rank and invertible. 

\section{Integrated Primary-Secondary Distribution Network Optimal Power Flow}
The integrated primary-secondary distribution network OPF problem can be formulated as follows:
\begin{subequations}
\begin{align}
     \min~~& Obj(PDNet)+Obj(SDNet)\\
     s.t.~~& \text{PDNet constraints}\\
     & \text{SDNet constraints}\\
     & \text{PDNet and SDNet coupling constraints}
\end{align}
\end{subequations}

For PDNet constraints, they have been widely investigated in existing works. The linearized PDNet power flow (\ref{eq:LPDNet}) \cite{LG} or the SDP-based PDNet power flow \cite{EDAHZ,LG} can both be applied to express PDNet constraints, which can be expressed as follows:
\begin{subequations}\label{eq:PDNetCon}
 \begin{align}
     \text{Linearized or SDP-based PDNet power flow}\\
      \bm{v}_I^{min}\leq\bm{v}_I\leq\bm{v}_I^{max}, \forall {n_I}\in\mathcal{N}
 \end{align}   
\end{subequations}
And SDNet constraints can be expressed as follows:
\begin{subequations}\label{eq:SDNetCon}
\begin{align}
   &\text{SOCP-ST or L-ST} ,  \forall{J*}\in\mathcal{J}\\
    &\text{L-TSL},  \forall{J*}\in\mathcal{J}\\
    \bm{v}&_i^{min}\leq\bm{v}_i^{J*}\leq\bm{v}_i^{max}, \forall {n_i^{J*}}\in\mathcal{N}^{J*}, \forall{J*}\in\mathcal{J}\\
    &\bm{p}_i^{J*}=\bm{p}_{i,l}^{J*}-\bm{p}_{i,d}^{J*}, \forall {n_i^{J*}}\in\mathcal{N}^{J*}, \forall{J*}\in\mathcal{J}\\
    &\bm{q}_i^{J*}=\bm{q}_{i,l}^{J*}-\bm{q}_{i,d}^{J*}, \forall {n_i^{J*}}\in\mathcal{N}^{J*}, \forall{J*}\in\mathcal{J}\\
    \nonumber&\bm{\underline{p}}_{i+}^{J*}\leq\bm{p}_{i+}^{J*}\leq\bm{\overline{p}}_{i+}^{J*}, \bm{\underline{p}}_{i-}^{J*}\leq\bm{p}_{i-}^{J*}\leq\bm{\overline{p}}_{i-}^{J*},\\ &\bm{\underline{p}}_{i+i-}^{J*}\leq\bm{p}_{i+i-}^{J*}\leq\bm{\overline{p}}_{i+i-}^{J*}, \forall {n_i^{J*}}\in\mathcal{N}^{J*}, \forall{J*}\in\mathcal{J}\\
    \nonumber&\bm{\underline{q}}_{i+}^{J*}\leq\bm{q}_{i+}^{J*}\leq\bm{\overline{q}}_{i+}^{J*}, \bm{\underline{q}}_{i-}^{J*}\leq\bm{q}_{i-}^{J*}\leq\bm{\overline{q}}_{i-}^{J*},\\ &\bm{\underline{q}}_{i+i-}^{J*}\leq\bm{q}_{i+i-}^{J*}\leq\bm{\overline{q}}_{i+i-}^{J*}, \forall {n_i^{J*}}\in\mathcal{N}^{J*}, \forall{J*}\in\mathcal{J}
\end{align}
\end{subequations}
where $\mathcal{J}$ denotes the node set that includes all coupling nodes between the PDNet and SDNets.  (\ref{eq:SDNetCon}a) is service transformer constraints,   (\ref{eq:SDNetCon}b)  is triplex service line constraints, (\ref{eq:SDNetCon}c) is SDNet voltage limits, (\ref{eq:SDNetCon}d)-(\ref{eq:SDNetCon}e) denote the values of net loads are equal to the values of loads minus the values of DERs, e.g., distributed generators,  (\ref{eq:SDNetCon}f)-(\ref{eq:SDNetCon}g) denote the power limits for net load consumption.

For the coupling constraints between the PDNet and any $SDNet(n_J,\phi_J)$, they are expressed as follows:
\begin{subequations}\label{eq:PSC}
\begin{align}
v_0^{J*}=v_{J}^{\phi_J}, \forall{J*}\in\mathcal{J}\\
P_{d_{0}}^{J*}+\text{j}Q_{d_{0}}^{J*}=p_{J}^{\phi_J}+\text{j}q_{J}^{\phi_J}, \forall{J*}\in\mathcal{J}
\end{align}
\end{subequations}
The equation (\ref{eq:PSC}a) indicates that the squared voltage magnitude $v_0^{J*}$ of node $n_0^{J*}$ in the $SDNet(n_J,\phi_J)$ is the same as the phase-$\phi_J$ squared voltage magnitude $v_J^{\phi_J}$ of bus $n_J$ in the PDNet. The equation (\ref{eq:PSC}b) indicates that the real and reactive power flows, $P_{d_0}^{J*}~\text{and}~Q_{d_0}^{J*}$, flowing through $z_{d_0}^{J*}$ are the same as the phase-$\phi_J$ real and reactive power loads of bus $n_J$ in the PDNet.

\section{Case Study}\label{sec:Case}
{In this section, we highlight the great significance of including SDNets in power flow analysis, and further evaluate the effectiveness and accuracy of our proposed models.}
\begin{figure*}[tbh]
    \centering
    \includegraphics[width=5.5in]{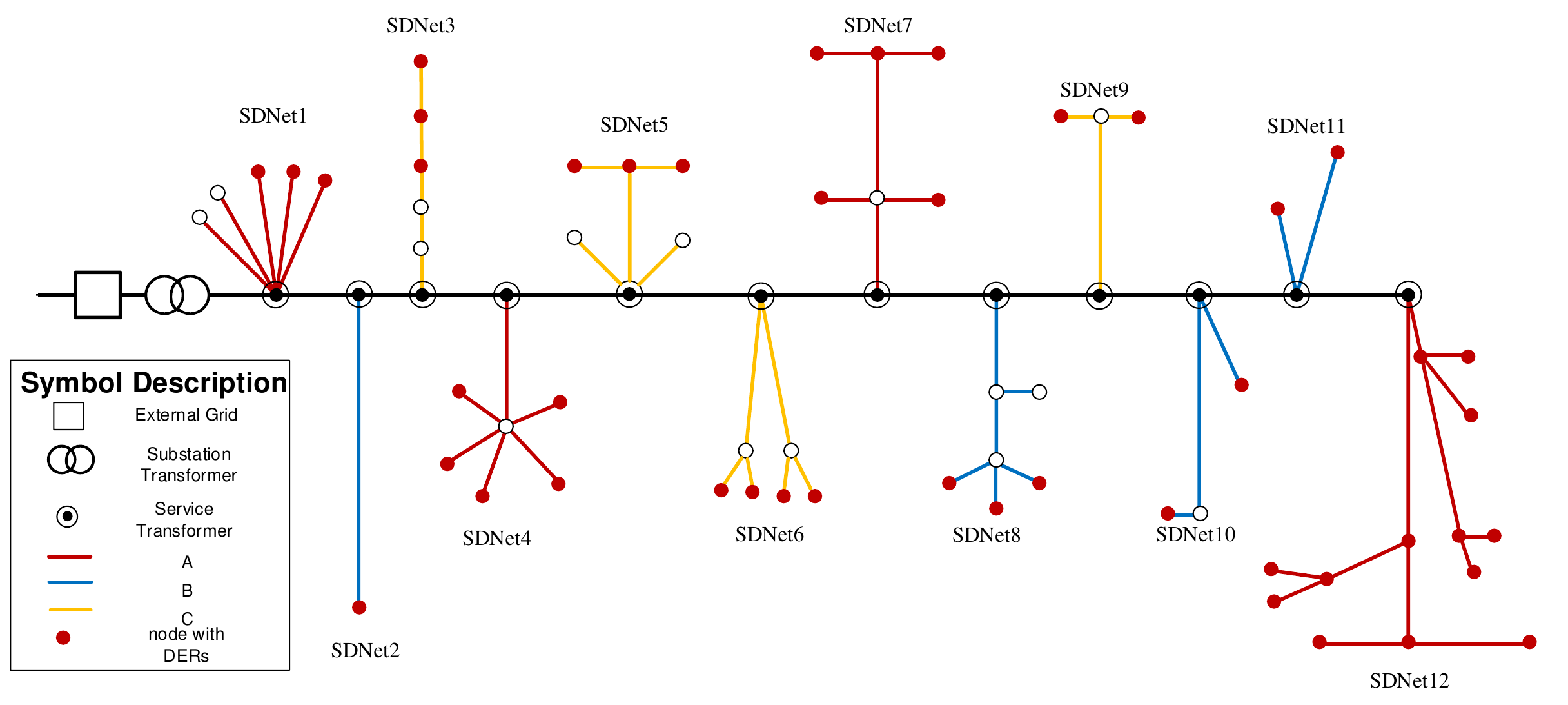}
    \caption{{A modified 12-bus PDNet system with 12 SDNets}}
    \label{fig:sdnet12}
\end{figure*}

\begin{figure}[tbh]
    \centering
    \includegraphics[width=3.3in]{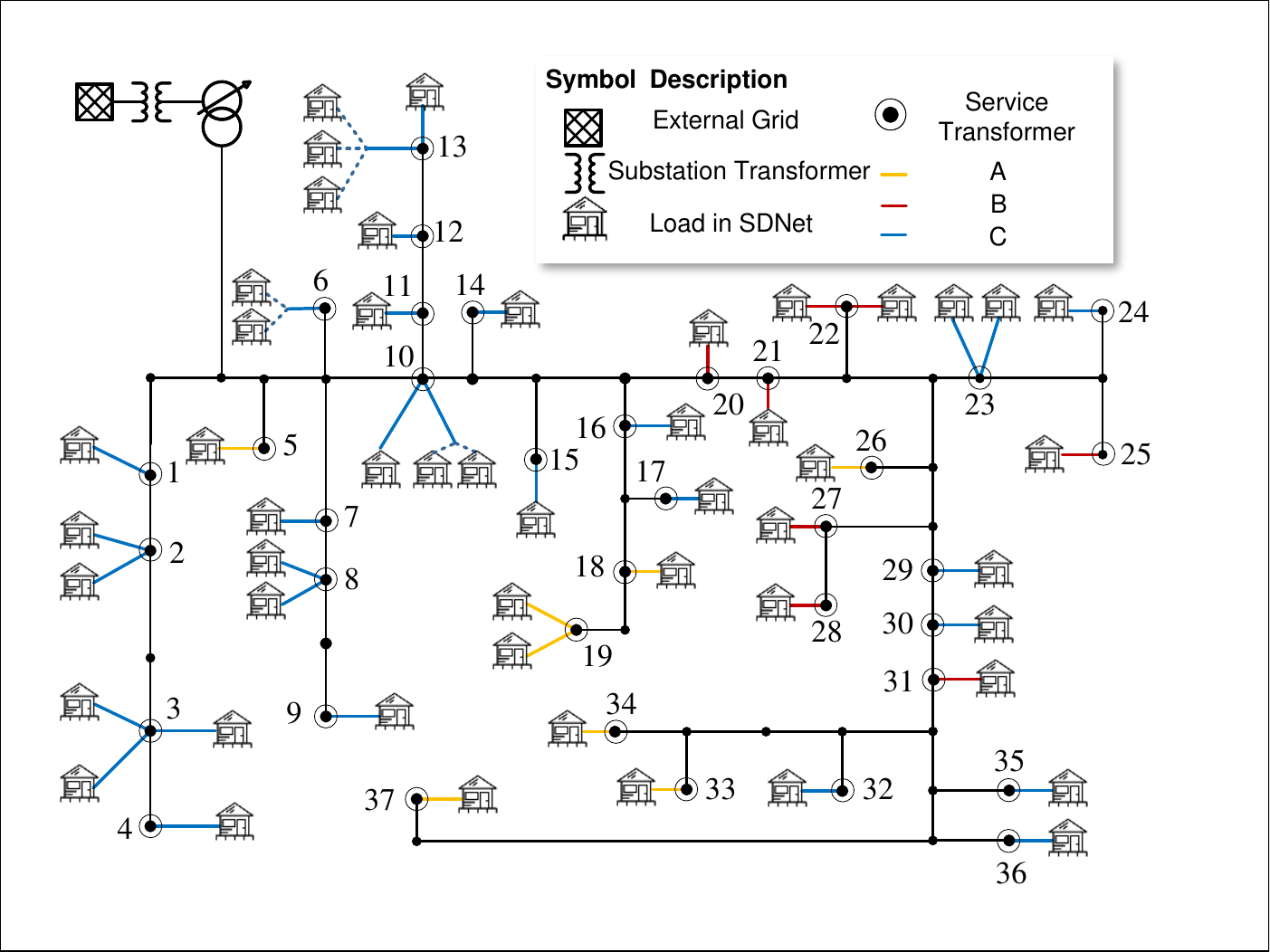}
    \caption{{A real utility primary-secondary distribution network in U.S.}}
    \label{fig:AMU}
\end{figure}
\subsection{Accuracy of L-ST and L-TSL}
To test the accuracy of L-ST and L-TSL, we consider the following three models:
\begin{itemize}
    \item Model 1 (PDNet model): This model only considers the PDNet model, where the loads and DERs in SDNets are aggregated at the PDNet bus. 
    {\item Model 2 (Accurate primary-secondary distribution networks): This model considers both PDNet and SDNet models, where SDNets are modeled by means of OpenDSS \cite{OpenDss} to consider the nonlinear SDNets power flows.
    \item  Model 3 (Simplified primary-secondary distribution networks): This model utilizes the L-ST and L-TSL to model SDNets.}
\end{itemize}

Specifically, Model 1 only considers the PDNet without SDNets. Model 2 considers both the PDNet and SNets, and performs the nonlinear power flow in primary-secondary distribution networks, which can be regarded as the benchmark. Model 3 also includes the PDNet and SDNets, but it performs the linearized PDNet power flow  in \cite{LG}
and the linearized SDNet power flow by our proposed L-ST and L-TSL.


\begin{figure}[thb]
    \centering 
    \includegraphics[width=3.0in]{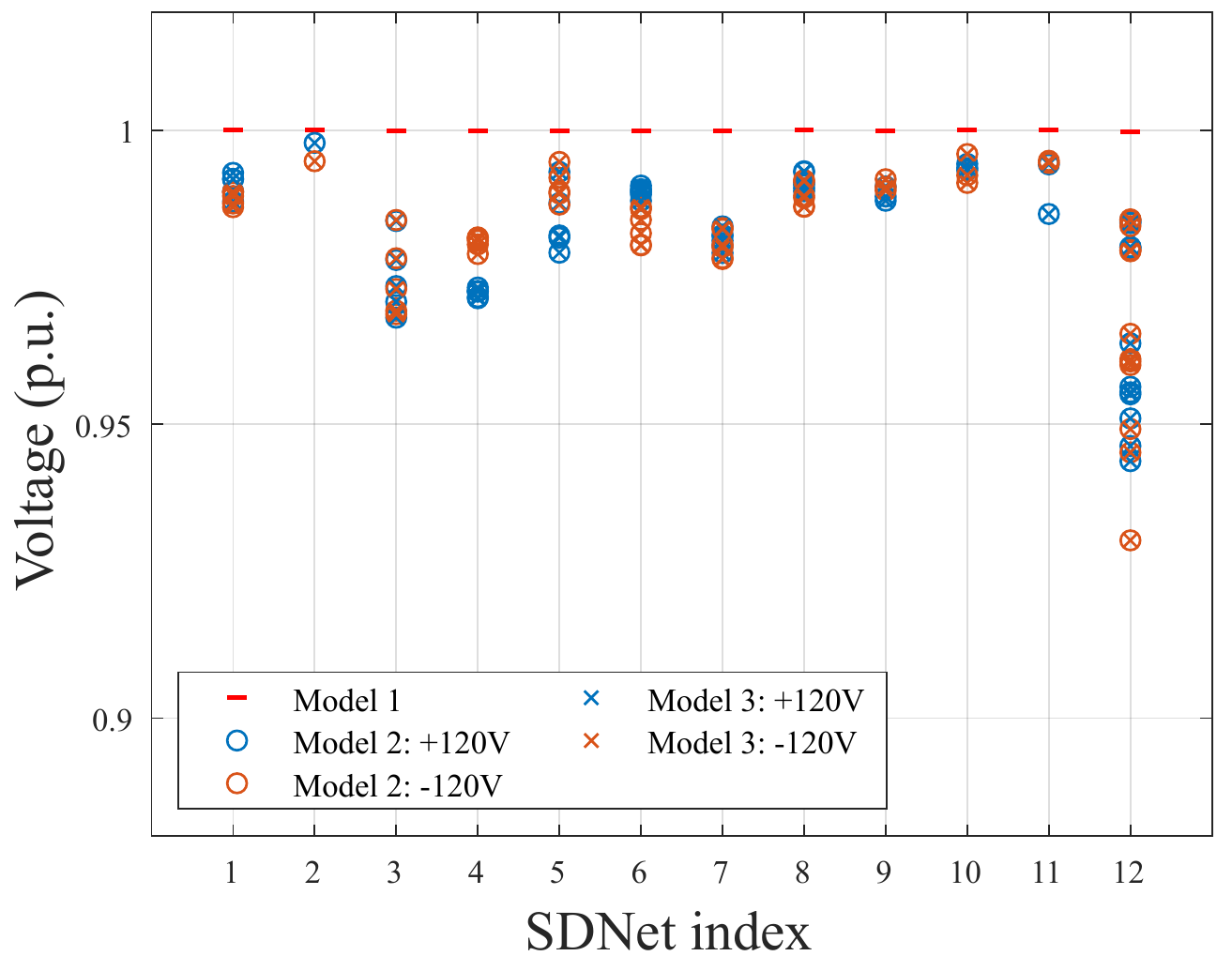}
    \caption{{Voltage distributions calculated by Models 1, 2, and 3 in the modified 12-bus PDNet with 12 SDNets.}}
    \label{fig:compare1&2_12bus}
\end{figure}
\begin{figure}[thb]
    \centering 
    \includegraphics[width=3.0in]{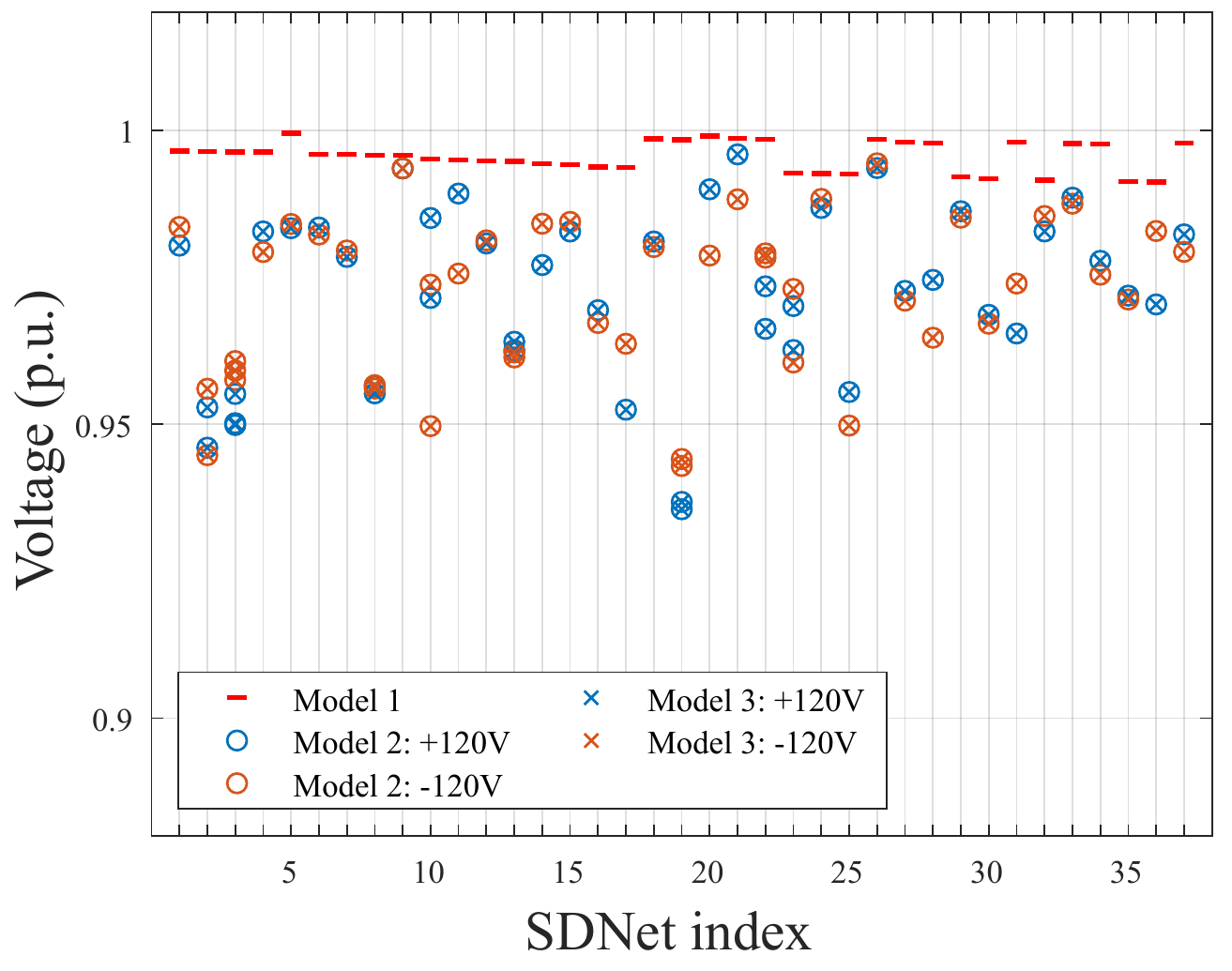}
    \caption{{Voltage distributions calculated by Models 1, 2, and 3 in the real utility primary-secondary distribution network in U.S.}}
    \label{fig:compare1&2_ec3}
\end{figure}

We first compare Models 1, 2 and 3 in two different test systems, one is a modified 12-bus PDNet system with 12 SDNets \cite{SDNet}-\cite{SDNetpara}, as shown in Fig.~\ref{fig:sdnet12}, another one is a real utility primary-secondary distribution network in the U.S., as shown in Fig.~\ref{fig:AMU}. Fig.~\ref{fig:compare1&2_12bus} and Fig.~\ref{fig:compare1&2_ec3} show voltage distributions calculated by Models 1, 2, and 3 in the above two test systems, respectively. As shown in Fig.~\ref{fig:compare1&2_12bus} and Fig.~\ref{fig:compare1&2_ec3}, Model 1  only reflects voltage distributions across the PDNets of two test systems, depicted by the red dotted line, due to the lack of SDNets. Instead, Models 2 and 3 exhibit the capability of depicting voltage distributions across each SDNet of two test systems, {where the voltage distributions for +120V and -120V circuits of SDNet are represented by the blue and orange markers,}  and the SDNet voltage distributions of Model 3 closely track Model 2. In addition, Models 2 and 3 can detect the under-voltage violations of SDNet 12 in the modified 12-bus PDNet system, shown in Fig.\ref{fig:compare1&2_12bus}, and SDNets 2 and 19 in the real  utility primary-secondary distribution network,  shown in  Fig.\ref{fig:compare1&2_ec3}, but  Model 1 fails to reflect the voltage violations in SDNets of test systems. The safe operation of PDNet cannot guarantee the safe operation of SDNet.

\begin{table}[htbp]\label{tab:12bus}
     \caption{Accuracy comparison: A modified 12-bus PDNet system with 12 SDNets}
    \centering
    \begin{tabular}{lccc}
        \toprule
         & P (\%) & $V_{\text{pri}}$ (p.u.) & $V_{\text{sec}}$ (p.u.)\\
        \hline
        Model1 & 1.88 &2.63e-06
  & -\\
        \hline
        Model3 & 1.89 & 2.55e-06
 & 5.69e-04
 \\
        \bottomrule
    \end{tabular}
\end{table}

\begin{table}[htbp]\label{tab:ec3}
    \caption{Accuracy comparison: A real utility primary-secondary distribution network }
    \centering
    \begin{tabular}{lccc}
        \toprule
         & P (\%) & $V_{\text{pri}}$ (p.u.) & $V_{\text{sec}}$ (p.u.)\\
        \hline
        Model1 & 2.02 & 2.09e-5  & -\\
        \hline
        Model3 & 2.47 & 2.21e-5 & 1.1e-3 \\
        \bottomrule
    \end{tabular}
\end{table}

Next, we consider a series of  different load conditions, and use  Model 2 to calculate the real power flows and voltage magnitudes as the benchmark. Also, we  utilize Models 1 and 3 to approximately estimate the power flows and voltage magnitudes, and compare their differences with Model 2, where the average errors of the power exchange at the substation transformer of PDNet, the PDNet voltage magnitudes, and the SDNet voltage magnitudes, i.e., $\text{P}(\%), V_{pri}(p.u.),V_{sec}(p.u.)$, are selected as the metrics to evaluate the performance of Models 1 and 3 compared to Model 2.  As shown in Tables I-II, the errors of $\text{P}(\%)$ for Models 1 and 3 are around 2$\%$ compared to Model 2 for both test systems. In addition, Model 1 cannot reflect the SDNet operations, which does not provide information for SDNet voltage magnitude errors $V_{sec}$. Instead, SDNet voltage magnitude errors $V_{sec}$ are very small of Model 3 compared to Model 2 in both test systems, which are 5.69e-04 and 1.1e-3, respectively.
The results indicate our proposed L-ST and L-TSL can provide a good estimate of real SDNet voltage magnitudes.

\subsection{Integrated Primary-Secondary Distribution Network OPF}
\begin{figure}[thb]
    \centering
    \includegraphics[width=3.0in]{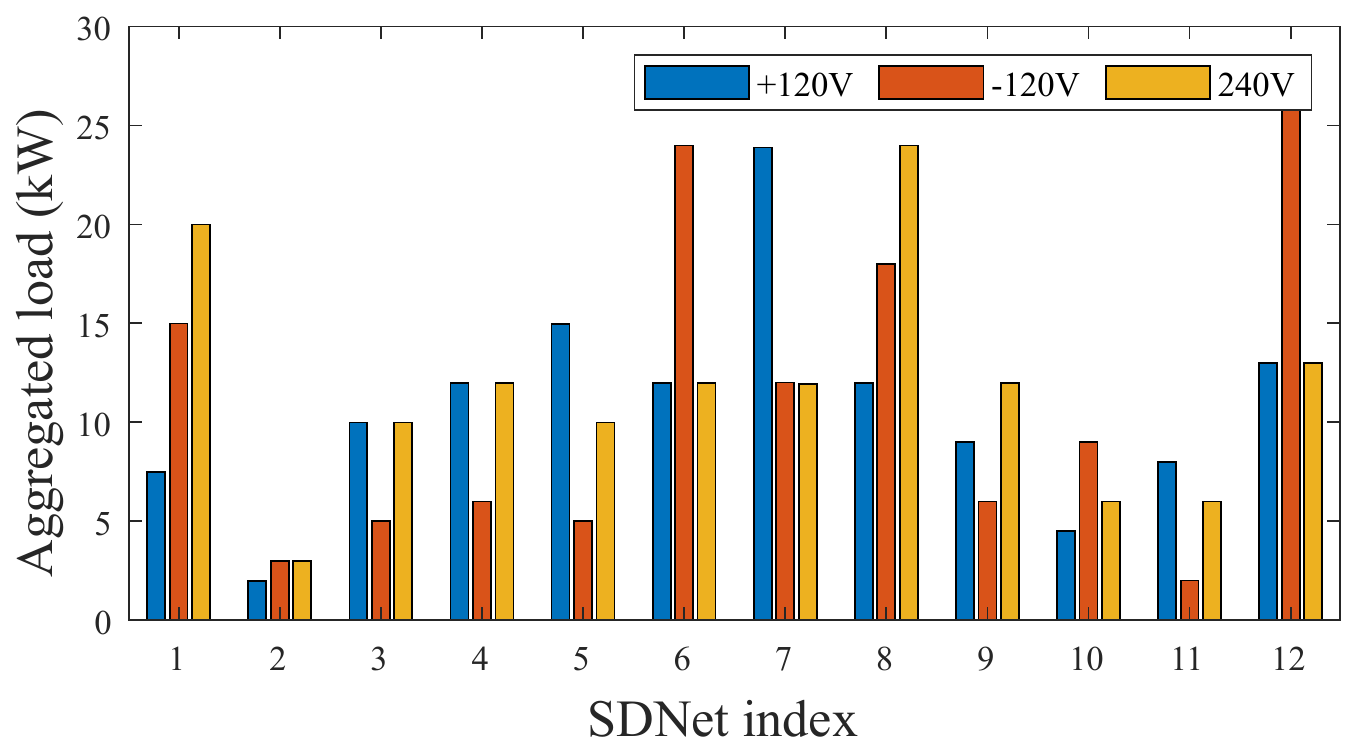}
    \caption{{{Aggregate load} consumption for each SDNet in the modified 12-bus PDNet with 12 SDNets}}
    \label{fig:SDNetPQ}
\end{figure}

This subsection considers a primary-secondary distribution network OPF problem, where the optimization goal is assumed to minimize SDNet power losses by dispatching DERs distributed across SDNets. And the linearized PDNet power flow, SOCP-ST and L-TSL are utilized in (\ref{eq:PDNetCon})-(\ref{eq:SDNetCon}). Taking the modified 12-bus PDNet system with 12 SDNets as one example, Fig.\ref{fig:SDNetPQ} depicts the {aggregate load} consumption for each SDNet. In addition, there are flexible DERs and loads, connected to SDNets shown in Fig.\ref{fig:sdnet12} and Fig.\ref{fig:AMU}. For these DERs distributed across SDNets, it is assumed that each DER can absorb/generate at most 1 kW real power and 1 kVar reactive power for the +120V circuit,  2 kW real power and 2 kVar reactive power for the -120V circuit, and 2 kW real power and 2 kVar reactive power for the 240V circuit. 

\begin{figure}[thb]
    \centering
    \includegraphics[width=3.0in]{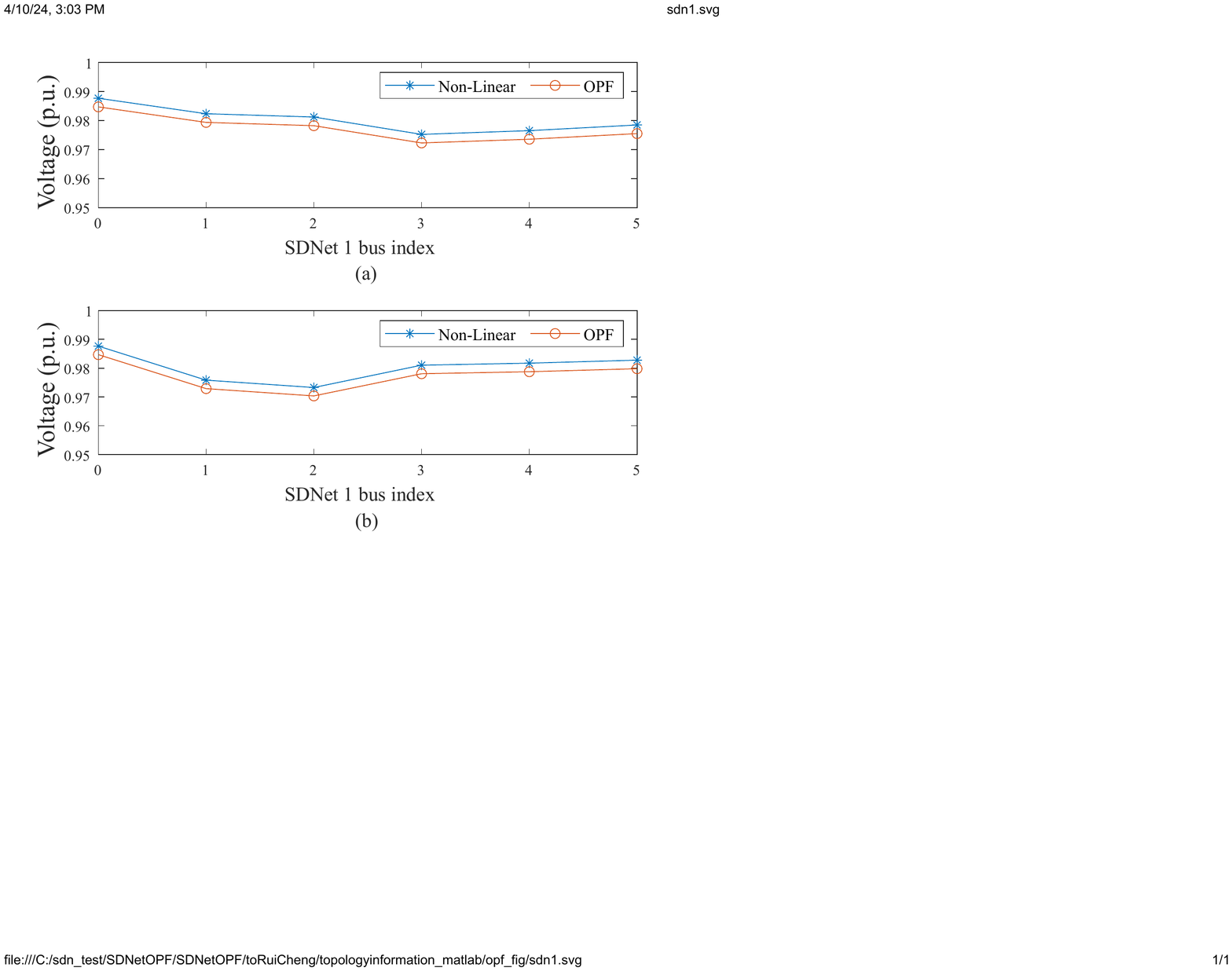}
    \caption{{Voltage magnitudes of SDNet 1: (a) +120V line. (b): -120V line.}}
    \label{fig:opfSDN1}
\end{figure}

\begin{figure}[thb]
    \centering
    \includegraphics[width=3.0in]{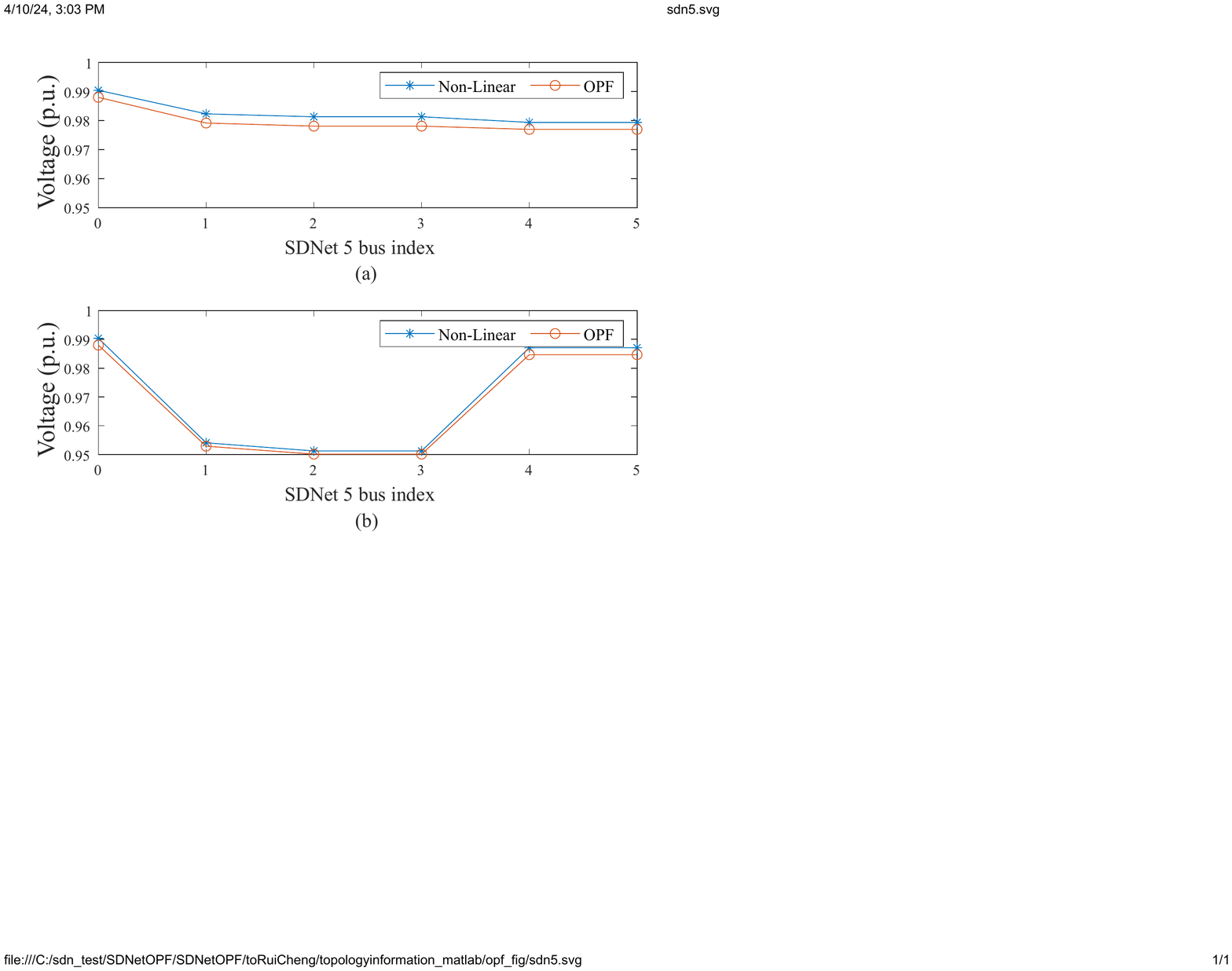}
    \caption{{Voltage magnitudes of SDNet 5: (a) +120V line. (b): -120V line.}}
    \label{fig:opfSDN5}
\end{figure}

Taking SDNets {1 and 5} as examples, the OPF-based voltage distributions, determined by this primary-secondary distribution system OPF, and the real voltage distributions, determined by inputting the optimal DER outputs into OpenDSS, are depicted in Fig.\ref{fig:opfSDN1} and Fig.\ref{fig:opfSDN5}. There are not many differences between the OPF-based
and real voltage distributions. This proposed integrated primary-secondary distribution network OPF effectively exhibits the capability of reflecting the voltage distributions and controlling DERs in SDNets.


\section{{Conclusions}}
This paper proposes an optimal power flow for integrated primary-secondary distribution networks with service transformers. Instead of neglecting SDNets with service transformers, we model SDNets, including service transformers and triplex service lines, in the integrated primary-secondary distribution network OPF. To be more specific, the SOCP relaxation and linearization for the service transformer power flow, i.e., SOCP-ST and L-ST, are proposed, and the linearized power flow for triplex service lines, i.e., L-TSL/ C-L-TSL,  is developed. Numerical studies are performed to show the effectiveness and superiority of our proposed models.


One remaining challenge is the observable information of SDNet models. For example, current sensing technologies only provide aggregate power demands at these customer locations, but do not independently provide the detailed information of loads at 120V and 240V levels. In the future, we will strive to resolve this disaggregation challenge. In addition, we will explore different power system applications in integrated primary-secondary distribution networks by means of our proposed integrated primary-secondary distribution network OPF.

\section*{Appendix A}
\noindent
\textit{Incidence Matrix Construction for Secondary Distribution Networks:}

\medskip

\begin{figure}[thb]
    \centering  \includegraphics[width=1.5in]{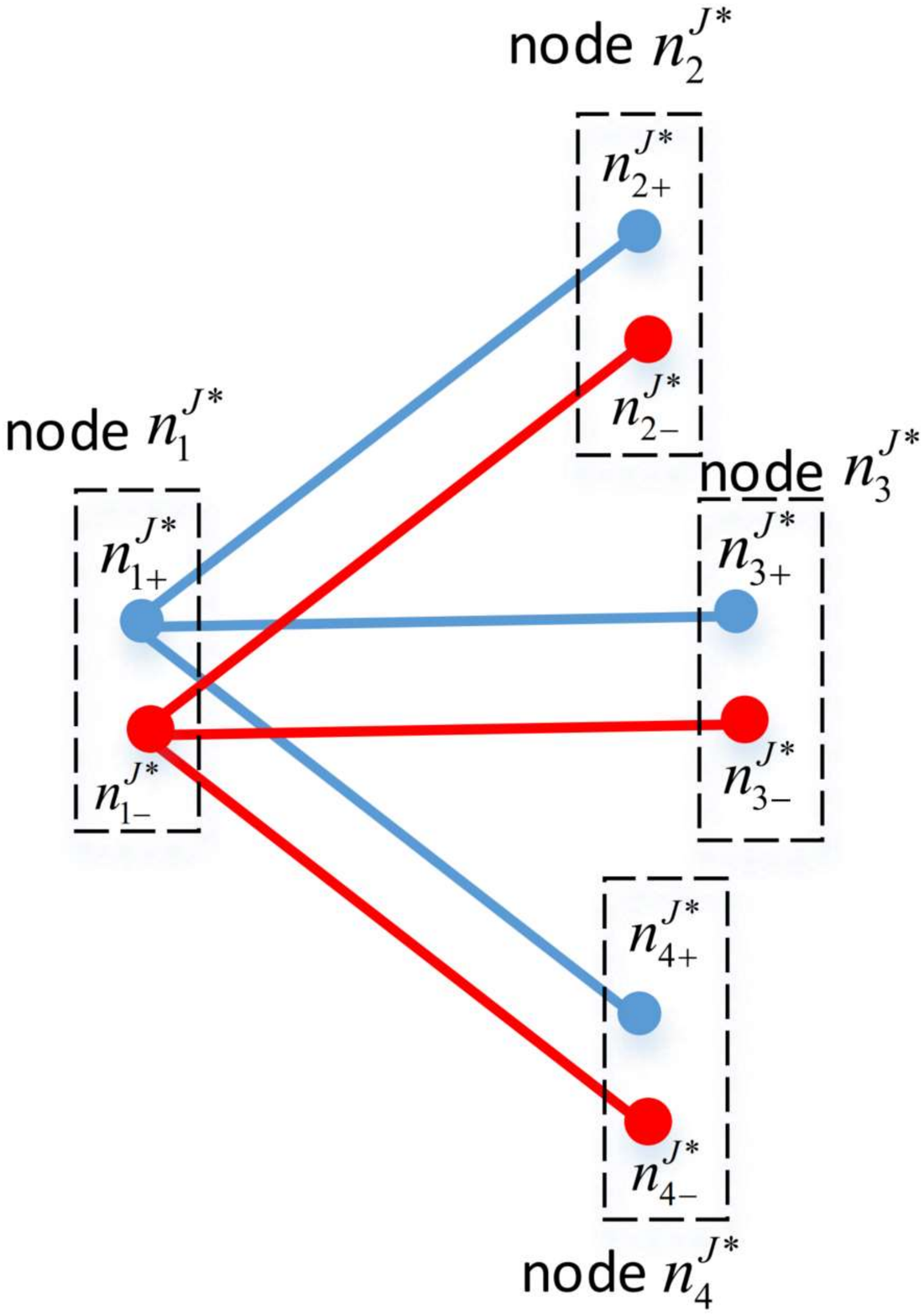}
    \caption{{Secondary distribution network illustration}}
    \label{fig:sdnetexample}
\end{figure}

To illustrate the construction of the incidence matrices $\bm{\bar{M}}^{J*}$ and $\bm{\bar{A}}^{J*}$ introduced in Section~\ref{sec:LTS} for triplex service lines in SDNets, consider Fig.\ref{fig:sdnetexample}.  This SDNet consists of 3 triplex service lines, where $\mathcal{L}^{J*}=\{(n_1^{J*},n_2^{J*}),(n_1^{J*},n_3^{J*}),(n_1^{J*},n_4^{J*})\}$.

Then, the incidence matrix $\bm{\bar{M}}^{J*}$ for +120V-wire circuit in this SDNet takes the following form:
\begin{equation}
    \bm{\bar{M}}^{J*}= \label{BarM}
    \begin{bmatrix}
    1&1&1\\
    -1&0&0\\
    0&-1&0\\
    0&0&-1
    \end{bmatrix}
\end{equation}
Next, the incidence matrix $\bar{\bm A}^{J*}={\bm{\bar{M}}}^{J*}\otimes{\textbf{I}_{2}}$ for the triplex service lines then takes the following form:
\begin{equation}
    \bm{\bar{A}}^{J*}=
    \begin{bmatrix}
        1&0&1&0&1&0\\
        0&1&0&1&0&1\\
        -1&0&0&0&0&0\\
        0&-1&0&0&0&0\\
        0&0&-1&0&0&0\\
        0&0&0&-1&0&0\\
        0&0&0&0&-1&0\\
        0&0&0&0&0&-1
    \end{bmatrix}
\end{equation}

\end{document}